\documentclass[a4paper,11pt]{article}
\usepackage{jheppub} 
\usepackage{cancel}
\usepackage{comment}
\usepackage{appendix}
\usepackage{orcidlink}
\usepackage[normalem]{ulem}

\title{Relativistic Dispersion Spectra across Lorentz boosted frames: Spurious modes and causality}

\author[a]{Sayantani Bhattacharyya,\orcidlink{0009-0001-4252-4705}}
\affiliation[a]{School of Mathematics and Maxwell Institute for Mathematical Sciences, University of Edinburgh, Peter Guthrie Tait Road, Edinburgh EH9 3FD, United Kingdom}

\author[b]{Sukanya Mitra,\orcidlink{0000-0002-2401-957X}}
\affiliation[b]{School of Physical Sciences, National Institute of Science Education and Research, An OCC of Homi Bhabha National Institute, Jatni-752050, India}

\author[c]{Shuvayu Roy,\orcidlink{0000-0002-5725-5712}}
\affiliation[c]{Department of Physics, Indian Institute of Technology, Gandhinagar, Gujarat 382055, India}

\author[d]{Rajeev Singh,\orcidlink{0000-0001-5855-4039}}
\affiliation[d]{Department of Physics, West University of Timisoara, Bulevardul Vasile P\^arvan 4, Timisoara 300223, Romania}

\emailAdd{sbhatta5@ed.ac.uk, sukanya.mitra@niser.ac.in, shuvayu.roy@iitgn.ac.in, rajeev.singh@e-uvt.ro}

\abstract{The analysis of excitation spectra in gradient-expanded relativistic fluid theories frequently leads to pathologies under Lorentz boosts. However, extracting the dispersion modes in a Lorentz boosted inertial frame can be nontrivial. Motivated by this problem, we develop a general framework for deriving the linearized dispersion spectra in Lorentz boosted frames using only information from the local rest-frame dispersion structure, particularly its mode-expansion coefficients. We observe that, under a Lorentz transformation from the local rest frame, additional boosted solutions may appear that conflict with the causality of the theory; we refer to these as ``spurious modes.” The key developments presented here are: (i) a convenient method for obtaining dispersion spectra across inertial frames, bypassing the traditional procedure of solving the boosted polynomial, and (ii) the establishment of a direct connection between mode conservation and the causality of a theory, supported by a detailed proof.}

\begin{document}
\maketitle
\flushbottom

\newpage

\section{Introduction}
\subsection{Motivation and theoretical background}
\label{sec:intro}
The study of linearized perturbations around stationary or equilibrium configurations provides a universal diagnostic for the theoretical admissibility of a broad class of physical systems via two benchmark criteria -- stability and causality~\cite{Hiscock:1987zz,Van:2007pw,Van:2008cy,Bu:2014sia,Bu:2015ika,Martinoia:2024hip,Roy:2023apk}. In fields like classical mechanics, gravitational physics, quantum field theory, condensed matter, and continuum dynamics, small perturbations around the equilibrium encode the dynamical response of the system and determine the nature of its excitations. These behaviors are described through the dispersion relation, obtained by linearizing the governing equations of motion of the state variables and seeking solutions for them by decomposing the perturbations into linearly independent modes. On a flat background manifold, at a spacetime point $(t,\vec{x})$, these perturbations are expressed in linearly independent Fourier modes of the form $e^{-i\omega t+ i \vec k \cdot \vec x}$, with $\omega$ and $\vec k$ as the frequency and wavenumber, respectively. The resulting dispersion polynomial in $(\omega,\vec k)$ from the linearized equations of motion, after solving, determines the allowed modes. The structure of the dispersion polynomial, as well as its roots, such as analyticity, stability, etc., then determines the physical properties of the theory under consideration~\cite{Olson:1990rzl,Heller:2020uuy,Grozdanov:2021jfw,Wang:2023csj,Boyanov:2024jge,Grozdanov:2025fwc,Geroch:1995bx,Arnold:2014jva}. 

A common occurrence found in many systems described in an effective-field-theory setup is the presence of pathologies upon adding higher-order derivative corrections to some background theory. In classical mechanics, the Ostrogradsky instability~\cite{Woodard:2015zca} is one such hurdle, while in quantum field theories, such higher-derivative corrections to the action can lead to problems with the unitarity and analyticity properties of the S-matrix \cite{Adams:2006sv}. In gravitational physics, as well as in relativistic hydrodynamics, one finds that such higher-order corrections lead to an ill-posed formulation of the initial value problem~\cite{Figueras:2024bba,Liu:2025xij}. The primary reason for these problems lies in the ill-posedness of the resulting equations of motion. A well-posed initial value formulation of such problems requires that the equations of motion governing the time evolution of the system be hyperbolic partial differential equations. Often, when these higher-derivative corrections are added to the theory, the governing equations of motion turn out to be parabolic. It has also been investigated that restoration of causality in a theory can be attempted by using infinite-order field redefinitions \cite{Mitra:2023ipl,Bhattacharyya:2023srn,Bhattacharyya:2024tfj}. Hence, the problem of diagnosing the presence of pathologies in a theory often translates into identifying issues with the nature of the partial differential equations of motion.

In the field of relativistic hydrodynamics, which is a low-energy effective theory, identical issues arise while describing the systems in terms of field perturbations of the conserved charges~\cite{Heinz:2013th,Romatschke:2017ejr,Andersson:2020phh}. As discussed in~\cite{Hiscock:1985zz,Denicol:2008ha,Pu:2009fj,Mitra:2021ubx}, in pathological theories, it is often difficult to diagnose the stability or causality-related issues in the local rest frame (LRF) of the fluid, since it is not the most general equilibrium state of a fluid. For this reason, analyzing such properties in an inertial frame where the fluid is moving against a static background makes these issues manifest in the case of a problematic theory of relativistic hydrodynamics. However, for a higher-order theory, a direct analysis of stability or causality properties by explicitly evaluating the roots of the dispersion polynomial in boosted frames can be extremely difficult and mathematically cumbersome \cite{Gavassino:2021owo}. Technically, the boost velocity $\vec{v}$ includes all possible scalar combinations of $\vec{k}$ and $\vec{v}$ for each $\omega$ power in a boosted polynomial, which makes the direct extraction of its roots a tedious exercise. Additionally, pathological theories in boosted frames can lead to unphysical modes, which were otherwise absent and typically diverge in the LRF of the fluid \cite{Kostadt:2000ty,Gavassino:2021kpi}. 
In this work, we will show that these additional boosted solutions represent the ``spurious modes", in the sense that their existence is connected to causality violation of the theory.
In recent work~\cite{Hoult:2023clg}, it has been indicated that these spurious modes lead to violating the large-$k$ causality criteria. Since these causality constraints at large-$k$ have been derived using the covariant stability condition at small-$k$ \cite{Heller:2022ejw}, these modes seem to violate causality criteria even in the small-$k$ limit.

In this work, we develop a general framework that connects the dispersion relations in two different inertial frames across a Lorentz transformation and predicts that the sole knowledge of the LRF dispersion mode (particularly the expansion coefficients) sufficiently provides us with the information to reconstruct the same mode in a boosted inertial frame. First, in Section~\ref{general}, we set up the framework and discuss the relation between the zeroes of the dispersion polynomial in two inertial frames of reference. We offer a parametric method for obtaining the spectrum of modes in a boosted frame from that of the LRF. Then, in Section \ref{nonspurious}, considering $\omega$ as an infinite series expansion in $k$, we set a one-to-one mapping between the expansion coefficients in the two frames for non-spurious modes. This finally provides us with the exact expression of a dispersion mode (both hydrodynamic and non-hydrodynamic) in a boosted frame, as long as the full mode information of the LRF is known. Thereafter, in Section \ref{spurious}, we demonstrate how the existence of spurious modes can be shown to be a consequence of a many-to-one mapping of the Lorentz transformation of the wavenumber on the complex-$k$ plane. Following this, we show that it either leads to the $\omega$ solution that includes poles or essential singularities in the small-$k$ domain, or shows a faster-than-linear $k$ growth (which is essentially superluminal) in the large-$k$ regime, both of which directly contradict causality. We summarize in Section \ref{summary} with some concluding remarks, future directions, and phenomenological implications. Appendix \ref{appen1} presents a proof of the uniqueness between a mode in the LRF and its counterpart in the boosted frame, which essentially reduces to the former as $v\rightarrow 0$. This proof is a crucial step in connecting the existence of spurious modes with violations of causality in our work.

Our investigation on the spurious modes and the violation of causality in Section~\ref{spurious} also complements the work in \cite{Hoult:2023clg}, which mentions that the number of modes for a causal relativistic theory must match in the small-$k$ and large-$k$ limits. Any increase in the number of roots at large momentum indicates a violation of mode conservation and, therefore, acausality. In our formulation, these extra roots correspond exactly to the boost-induced spurious modes seen in the transformed dispersion polynomial. Thus, our work reinforces the claim that the appearance of spurious modes provides a direct, physically interpretable signal of causality breakdown in relativistic hydrodynamics and, more broadly, in any linearized effective theory. The framework developed here offers a general algorithm to calculate such unphysical excitations.
\subsection{Notation and convention}
\label{notcon}
We define $\tilde{\omega}$ and $\tilde{k}$ as the frequency and wavenumber in the local rest frame (Lorentz frame where the fluid is static). We denote by $\omega$ and $k$ the corresponding quantities in the boosted frame, which moves with a uniform background velocity $v$ relative to the LRF. The hydrodynamic modes of the boosted spectra are denoted separately by $\omega_q$ in Section~\ref{subsec:formal_general_formulae}. Apart from this, any quantity represented by $X$ in the LRF carries a subscript $v$ in the Lorentz boosted frame, written as $X_v(v)$. The argument $(v)$ indicates the explicit dependence of the quantity on the boost velocity $v$. Following these conventions, the functions $P(\tilde{\omega},\tilde{k})$ and $P_v(\omega,k,v)$ denote the dispersion polynomials in the LRF and in the boosted frame, respectively. Analogously, we denote the roots (zeroes) of the dispersion polynomial by $W(\tilde{k})$ for the LRF and $W_v(v,k)$ for the boosted frame. In Section~\ref{subsec:mapping}, we have introduced a parameter $p$ to find a mapping between the roots in LRF and boosted frame. The coefficients $P_n$ and $B_n$ introduced in Section~\ref{subsec:mapping} denote the expansion coefficients of $p$ in terms of $k$ and $v$.

Next, we mention here a number of coefficients used in Section~\ref{nonspurious}. $a_n$ are the coefficients of the LRF mode expanded around zero momentum. The quantities $\tilde{a}_n$ and 
$a^*_n$\footnote{The notation $a^*_n$ must not be confused with the complex conjugate of $a_n$.} denote the corresponding 
coefficients for the hydrodynamic and non-hydrodynamic modes, respectively, in the boosted reference frame. The coefficients ${\cal{A}}_n,{\cal{B}}_n,b_n,c_n,d_n,Q_n,\alpha_n,\alpha_n^*$ are introduced in Section \ref{nonspurious} to facilitate the intermediate steps of extracting the boosted expansion coefficients from the LRF ones and present the results in a precise manner.

Throughout this work, we adopt natural units $\hbar = c = k_B = 1$ and use a mostly positive metric signature $g^{\mu\nu} = \{-1,1,1,1\}$.

\section{Spectrum in the boosted frame}
\label{general}
The spectrum of linearized perturbations around any equilibrium or `ground state' configuration (ones that do not evolve with time) plays a very important role in understanding the properties of the equilibrium and the entire system in general. In the case of hydrodynamics, a uniformly moving fluid at constant temperature constitutes an equilibrium configuration in flat space. It is well-known that the spectrum of linearized perturbations, including the number of allowed modes, drastically changes as one moves from `zero velocity' fluids to uniformly moving fluids. However, these two equilibrium fluid configurations (i.e., static fluids and fluids with uniform velocity) are related by a Lorentz transformation. Therefore, whenever the underlying hydrodynamic theory is Lorentz-invariant, we should be able to predict the spectrum around the moving fluid from that of the static fluid and vice-versa, using some form of Lorentz transformation applied in a mode-by-mode fashion.

In this section, we shall provide a set of equations that will inform us how to determine the mode(s) in the boosted frame for a given fluid from every mode in the LRF.
\subsection{Theoretical set-up}

In this subsection, we would like to investigate how the dispersion polynomial for a given fluid, derived with the static background, is related to that derived in a frame
where the background is moving with a certain nonzero velocity $\vec v$.

Suppose the background fluid is moving with a uniform velocity $\vec v$. The 4-velocity $u^\mu$ for the perturbed fluid  will have the structure as
\begin{align}
u^\mu =v^\mu + \delta u^\mu~ e^{-i\omega t +i\vec k \cdot \vec x} =v^\mu + \delta u^\mu~ e^{-i\omega t +ik_x x + i k_y y},~~~~v^\mu =\gamma\{1,-v,0,0\}~,
\label{frame1}
\end{align}
with $\gamma= {\frac{1}{\sqrt{1-v^2}}}$ being the Lorentz boost factor.
Here, we have chosen the direction of the background fluid velocity to be the $x$ direction. The component of the spatial momenta along the boost direction is denoted as $k_x$. The $y$ direction is chosen along the projection of the spatial momentum in the direction perpendicular to the background fluid velocity.

Now we apply the following boost transformation between the moving frame given in \eqref{frame1} and the LRF,
\begin{equation}\label{boost}
\begin{split}
t\rightarrow \tilde t = \gamma ( t + v x),~~~x\rightarrow\tilde x = \gamma (x + vt),~~~y\rightarrow\tilde 
y =y~.
\end{split}
\end{equation}
Under this transformation, we have $v^\mu \rightarrow\tilde{ v}^\mu = \{1,0,0,0\}$ which means that the background fluid is static. The combination $\left[-\omega t + k_x x + k_y y \right]$ appearing in the fluctuating part of Eq.~\eqref{frame1} transforms as,
\begin{align}
\left[-\omega t + k_x x + k_y y\right]=& -\gamma \omega (\tilde t - v\tilde x) +\gamma k_x(\tilde x - v\tilde t) + k_y\tilde y\,,\nonumber\\
=&-\gamma(\omega + v k_x) \tilde t + \gamma (k_x + v\omega)\tilde x + k_y\tilde y  \,.
\label{frame2}
\end{align}
Consequently, in the new frame, the fluid velocity takes the following form, 
\begin{equation}\label{pert}
\begin{split}
&\tilde u^\mu = \tilde v^\mu +\delta \tilde u^\mu ~e^{-i\tilde\omega \tilde t + i \tilde k_x \tilde x + \tilde k_y\tilde y}~,\\
{\text{with,}}~~~&\tilde \omega = \gamma(\omega + v k_x) ,~~~\tilde k_x =\gamma(k_x+ v\omega),~~\tilde k_y = k_y~.
\end{split}
\end{equation}
As expected in tilde coordinates, the perturbation around a moving fluid takes the identical structure to that of a static fluid. Particularly, the functional form of the dispersion polynomial in terms of $\{\tilde\omega,\tilde k\}$ is exactly the same as that of the standard dispersion polynomial one computes around a static fluid.

In summary,  the overall transformation of the frequency $\omega$ and the wave-vector $\vec k$ under boost $v$, (in the same direction as that of the wave-vector $\vec k$) has the well-known formula,
\begin{equation}\label{boost1}
\begin{split}
\omega\rightarrow\tilde\omega&= \gamma\left(\omega + v ~k\right)\,,\\
k\rightarrow\tilde k &= \gamma\left(k +v~\omega\right),~~~\text{with}~~k = \sqrt {\vec k\cdot\vec k}~.
\end{split}
\end{equation} 
From now on, for analytical simplicity, we are considering the case with $k_y=0$.

By calculating the equations of motion for a given theory, we can derive the dispersion polynomial consisting of the frequency and the wavenumber of the linearized perturbations. As mentioned before, we define the dispersion polynomial in the LRF as $P(\tilde \omega, \tilde k)$, whereas, in the boosted frame, it is defined as $P_v (\omega, k, v),$ which has an explicit dependence on the boost parameter $v$. Following this prescription and
applying the transformation given in Eq.~\eqref{boost1} in $P(\tilde \omega, \tilde k)$, the dispersion polynomial in the moving frame becomes,
\begin{equation}\label{dispoltrans}
    P(\tilde \omega, \tilde k)\rightarrow P_v(\omega,k,v) = P (\gamma (\omega + v~k), \gamma(k + v~ \omega))~.
\end{equation}
The zeroes of the dispersion polynomials give us the dispersion relations, i.e., the wavenumber dependent functional forms of the frequency of the allowed perturbation modes. Thus, the zeroes of $P(\tilde \omega, \tilde k)$ give us the $\tilde \omega (\tilde k)$ dispersion relations in the LRF, and the zeroes of $P_v(\omega, k, v)$ give us  $\omega(k,v)$ spectrum in the boosted frame.

\textbf{Note:} For clarity, we would like to elaborate on what we mean by `boosted frame' in the subsequent sections. A `boosted frame' in our discussion would mean any frame Lorentz boosted with respect to the LRF, one in which the background fluid is in motion. 


\subsection{The modes in the boosted frame from the local rest frame modes}
\label{subsec:mapping}

In this subsection, our aim is to find the relation between the boosted frame modes and the LRF modes. In other words, we would like to find the relation between the zeroes of the two polynomials $P_v(\omega, k,v)$ and $P(\tilde \omega, \tilde k)$.
 
Let $\tilde \omega = W^i(\tilde k)$, where $i\in\{1,2,\cdots, M\}$, be the $M$ number of zeroes of $P(\tilde \omega, \tilde k)$ that allows us to write the polynomials in the LRF and the boosted frame, respectively, as
\begin{align}\label{boostfactor}
P(\tilde\omega,\tilde k) &=\prod_{i=1}^M\left[\tilde\omega - W^i(\tilde k)\right]\nonumber\\
P_v(\omega,k,v) &= \prod_{i=1}^M\left[\gamma(\omega+v k) - W^i(\gamma(k+v\omega))\right]\equiv \prod_{i=1}^M S_i(\omega,k)~,
\end{align}
where, for convenience, we introduce the quantity $S_i(\omega,k) \equiv \left[\gamma(\omega+v k) - W^i(\gamma(k+v\omega))\right]$.

Whenever $P_v(\omega,k,v)$ vanishes, at least one of the factors $S_i(\omega,k)$ must also vanish. In other words, the zeroes of $P_v(\omega,k,v)$ must be a zero of at least one of $S_i(\omega,k)$.
Now the zeroes of $S_i(\omega,k)$ could have a simple parametric solution as,
\begin{equation}\label{solpara2}
k = \gamma\left\{p-v~W^i(p)\right\},~~~\omega = \gamma\left\{W^i(p)-v~p\right\}~.
\end{equation}
Equation~\eqref{solpara2}\footnote{
We note that the free parameter $p$ coincides with the value of the boosted wavenumber $\tilde k$, as can be shown by inverting equation \eqref{solpara2}, and comparing the results with \eqref{boost1}. However, we keep the notation $p$ here to emphasize its parametric role.
}
is a parametric solution in the sense that if it is substituted into $S_i(\omega, k)$, it identically vanishes for every $p$. This can be seen easily as follows,
\begin{align}
S_i(\omega,k) &= \gamma(\omega + v~k) - W^i (\gamma(\omega+v~k))~,\nonumber \\
    &= \gamma^2\left[ W^i(p) - v~p ~+ v\{p ~- v~W^i(p)\}  \right] - W^i \left[ \gamma^2\left\{ p - v~W^i(p)~+ v (W^i(p) - v~p) \right\} \right]~,\nonumber \\
    &=W^i(p) - W^i(p) = 0~,~~~~\text{for all values of} ~~p~.
   \end{align}
Note that by varying $p$, we can access all values of $k$ in the complex $k$-plane and the corresponding value for $\omega$. Therefore, this parametric solution will certainly generate one zero for the $P_v(\omega,k,v)$ from each $S_i$ and thus at least $M$ zeroes of $P_v(\omega,k,v)$ could be constructed in one-to-one correspondence with the $M$ modes in LRF.

Now let us explore the mechanism of generating extra modes in the boosted frame that cannot be mapped to any LRF mode in the $v\rightarrow 0$ limit.

Suppose, for a certain $i=i_{0}$, the LRF mode $W^{i_{0}}(p)\equiv {\mathcal W}(p)$ is such that the first equation given in \eqref{solpara2}
has more than one solution for $p$, for every complex $k$~\footnote{LRF modes - $W^i(p)$ are typically not polynomials in $p$, and therefore the multiple solutions are not always guaranteed for every $i$.}. 
Let $p_a(k)$, where $a\in\{1,2,\cdots,n\}$, be $n$ such distinct solutions for a fixed $k$. In other words,
\begin{equation}\label{defa}
p_a(k)\neq p_b(k)~~\text{if}~~a\neq b,~~\text{but}~~\gamma\left\{p_a(k)-v~{\cal W}(p_a(k))\right\}  =k \quad \forall \,\, a\in\{1,2,\cdots,n\}.
\end{equation}
Substituting them into the expression for $\omega$ in the second equation of~\eqref{solpara2} we find the expression for some of the zeroes of $P_v(\omega,k,v)$ as the following,
\begin{equation}\label{boostsol}
\omega_a(k) \equiv \omega(p_a(k)) = \left(\frac{1}{v}\right)\left[{\frac{p_a(k)}{\gamma}} -k\right],~~a =\{1,2,\cdots,n\}~~\Rightarrow~~\omega_a(k) \neq\omega_b(k)~~\text{if}~~a\neq b\,.
\end{equation}
Equation~\eqref{boostsol} means that, for a single value of $k$ we have multiple distinct values of $\omega(\equiv\omega_a)$ that simultaneously solve both the equations in \eqref{solpara2} and hence give $S_{i_{0}}(\omega,k)=0$, and generate modes or zeroes for $P_v(\omega,k,v)$.

Now, if the solutions, $p_1, p_2,\cdots, p_n$,  exist for every complex $k$ and real $v<1$, then from the perspective of the boosted frame, there will be $n$ distinct zeroes of $P_v(\omega,k,v)$, though generated from a single factor $S_{i_0}$ or a single LRF mode ${\cal W}(\tilde k) \equiv W^{i_{0}}(\tilde k)$. In other words, the existence of such multiple solutions indicates the presence of spurious modes.

Note that one could always generate all the zeroes (both spurious or non-spurious) of the boosted polynomial $P_v(\omega,k,v)$ from the LRF modes by eliminating $p$ from the relations given in \eqref{solpara2}. In this sense, it is an alternative way to find the modes in a boosted frame.
However, we would also like to emphasize that, in practice, the elimination of $p$ from the relations \eqref{solpara2} might not be simpler than solving for the zeroes of $P_v(\omega,k,v)$ directly. The main advantage of this parametric solution for the zeroes of $P_v(\omega,k,v)$ is that it relates the spectrum around the moving fluid to the LRF modes in a one-to-one or many-to-one (when there exist spurious modes) fashion. This gives us a clear prescription of how to translate any physical constraints on the LRF spectrum to any boosted frame. For example, it provides a connection between the causal LRF spectrum and the existence of spurious modes, as we shall see in Section~\ref{spurious}~\footnote{Reference~\cite{Hoult:2023clg} also discusses the connection between the acausality of the theory and the spurious modes. However, our analysis provides another way of looking at the same problem.}.

In the rest of this subsection, we shall present a perturbative way of eliminating $p$ from \eqref{solpara2} around $k=0$. This is the regime of spatial momentum that we expect to be covered by hydrodynamics in any inertial frame, particularly the boosted ones. 

Let us choose the generic LRF mode $W^{i_{0}}(p)={\cal W}(p)$ which has multiple solutions for the first equation of \eqref{solpara2} for every complex $k$ and therefore $k=0$.
Suppose $p = k_s(v)$ is one of the multiple zeroes of the first equation in \eqref{solpara2}. In other words, $k_s(v)$ satisfies,
 \begin{equation}\label{kss}
 k_s -v~{\cal W}(k_s)=0~.
 \end{equation}
 We then perform a Taylor expansion of ${\cal W}(p)$ around $p = k_s$ in the following manner,
 \begin{align}
 {\cal W}(p) = \sum_{n=0}^{\infty}\left[{\cal W}^{(n)}\over n!\right] ( p-k_s)^n~,~~~\text{where}~~~{\cal W}^{(n)}\equiv \left[d^n {\cal W}(p)\over dp^n\right]_{p=k_s}~.
 \label{roots}
 \end{align}
 We next substitute the following ansatz for $p(k,v)$ in the first equation of \eqref{solpara2},
 \begin{align}\label{ansatz01}
     p (k,v) =k_s(v)+ \sum_{n=1}^\infty P_n(v)\left(k\over \gamma\right)^n~,
      \end{align}
 which for $k=0$ reduces to $p=k_s(v)$. After the substitution, it follows that
\begin{align}\label{ansatz1}
{k\over\gamma}~& =~ k_s +  \sum_{n=1}^\infty P_n(v)\left(k\over \gamma\right)^n-v~{\cal W}\bigg( k_s +  \sum_{n=1}^\infty P_n(v)\left(k\over \gamma\right)^n\bigg)~,\nonumber\\
& =~ \sum_{n=1}^\infty P_n(v)\left(k\over \gamma\right)^n - v ~{\cal W}^{(1)}\sum_{n=1}^\infty P_n(v)\left(k\over \gamma\right)^n\nonumber\\&~~~~~~~~~~~- \left[v ~{\cal W}^{(2)}\over 2\right]\sum_{m,n=1}^\infty P_n(v)~P_m(v)\left(k\over \gamma\right)^{m+n}+\cdots~.
\end{align}
Knowing the coefficients $P_n$ as a function of $v$ amounts to the inversion of the relation between $k$ and $p$.

Next, comparing the powers of $\left(k\over\gamma\right) $ on both sides of the Eq.~\eqref{ansatz1}, we obtain the solutions for $P_n(v)$ in terms of ${\cal W}^{(n)}$. The first three coefficients can be calculated to be
\begin{align}\label{solp1}
    &P_1(v) = {1\over 1- v~{\cal W}^{(1)}}~,\qquad P_2(v) = {v~{\cal W}^{(2)}\over 2\left\{1- v~{\cal W}_i^{(1)}\right\}^3}~,\nonumber\\
    &P_3(v) = {\left(v~{\cal W}^{(2)}\right)^2\over 2\left\{1- v~{\cal W}^{(1)}\right\}^5} +{v~{\cal W}^{(3)}\over 6\left\{1- v~{\cal W}^{(1)}\right\}^4}~.
\end{align}
Once we know $p$ in terms of $k$ (from Eq.~\eqref{ansatz01}) as an expansion around $k=0$ and derive the $P_n$ coefficients, we substitute it into the second equation of \eqref{solpara2} and find the frequency $\omega(k)$ in the boosted frame in a power series around $k=0$ as\footnote{Ideally the expression of  $\omega(k)$ - the mode in the boosted frame, derived using the parametric solution \eqref{solpara2} and elimination of the parameter $p$, should carry two further indices. One of them would indicate that this boosted mode is generated from the $i_0^{th}$ LRF mode, i.e., we substituted $W^{i_{0}}(p) \equiv {\cal W}(p)$ in the first equation of \eqref{solpara2}. The second one should indicate that it is a solution around a particular zero $p= k_s(v)$  among the multiple zeroes of the first equation of \eqref{solpara2}, i.e., we have chosen a particular value for the $a$ index, introduced in equation \eqref{defa}. However, to keep the notation uncluttered, we have chosen to suppress all these indices.},
\begin{equation}\label{pertomega}
\omega(k) = \left(1\over v\right)\left\{{p(k)\over\gamma} - k\right\} =\left(1\over v\right)\left[{k_s(v)+\sum_{n=1}^\infty P_n(v)\left(k\over \gamma\right)^n\over\gamma} - k\right] ~.
\end{equation}

A couple of points to note here:
\begin{itemize}
\item $\omega(k)$ diverges in the $v\rightarrow 0$ limit, unless $\left[\lim_{v\rightarrow 0}k_s(v)\right]=0$. 

It could easily be seen by observing Eq.~\eqref{solp1} that $\lim_{v\rightarrow0}P_n(v) = \delta_{n,1}$. Therefore, it follows that,
\begin{align}\label{zeroomega}
\lim_{v\rightarrow0}\omega(k) &=\lim_{v\rightarrow0}\left[{1\over v}\left\{{k_s(v)+\sum_{n=1}^\infty P_n(v)\left(k\over \gamma\right)^n\over\gamma} - k\right\} \right]= \lim_{v\rightarrow0}\bigg[{k_s(v)\over v}\bigg]~,
\end{align}
where we have also used the fact that $\lim_{v\rightarrow 0}\gamma =1$. In other words, whenever the equation \eqref{kss} has a finite solution even in the limit of vanishing boost (i.e., $v\rightarrow0$ limit), it corresponds to a mode in the boosted frame that does not exist in LRF, and therefore it is a spurious mode.
\item Now let's consider the case where ${\cal W}(p)$ is itself a  hydrodynamic mode in LRF. These are the modes that satisfy,
$$\lim_{p\rightarrow 0}{\cal W}(p)=0~.$$
For every such mode in the LRF, $k_s=0$ always satisfies the equation \eqref{kss} for every $v$. From the previous arguments given around Eq.~\eqref{pertomega} and \eqref{zeroomega}, we observe that they will generate the non-spurious hydrodynamic modes in the boosted frame. 
In other words, for hydrodynamic modes, the boost transformation maps the region near the origin of the complex $p$ plane in LRF to the neighbourhood of the origin of the complex $k$ plane in the boosted frame. 
\item Further, we could see that (see Appendix \ref{appen1} for details) if there are $M$ modes in the LRF, then there would be exactly $M$ modes in the boosted frame that smoothly merge to the $M$ LRF modes in the $v\rightarrow 0$ limit. 
In particular, a hydrodynamic mode in the boosted frame will always map back to a hydrodynamic mode in the LRF.
 
\item For the non-hydrodynamic modes in LRF, $k_s(v)$ is finite for finite $v$ but it approaches zero as $v\rightarrow 0$. Here, the region near the neighbourhood of $k_s(v)$ - a point at a finite, and $v$ dependent distance away from the origin of the $p$ plane, maps to the neighbourhood of the origin of the $k$ plane in the boosted frame. 

The exact expression for $k_s(v)$ always depends on the exact form of the ${\cal W}(p)$. 
In hydrodynamic theories, which are treated in derivative expansion,  all dispersion relations, i.e., all LRF modes - ${\cal W}(p)$ (including the non-hydrodynamic ones)  are typically expressed in an expansion around $p=0$. Such an expansion could be used to determine $k_s(v)$ even for non-hydrodynamic LRF modes,  provided it is within the radius of convergence of the series~\cite{Bhattacharyya:2024ohn}.

Now, as the distance between the point $p = k_s(v)$ and the origin of the $p$ plane gradually decreases with $v\rightarrow 0$, there always exists a sufficiently small $v$ for which $k_s(v)$ will be within this radius of convergence. 

In other words, for non-hydrodynamic modes in LRF, it is possible to determine a perturbative expansion of  $k_s(v)$ in powers of small $v$.
\begin{itemize}
\item We assume the following ansatz for $k_s(v)$,
\begin{equation}
  k_s(v) = \sum_{n=1}B_n~v^n~.  
  \label{ksexpansion}
\end{equation}
\item We substitute this ansatz in equation \eqref{kss} and expand ${\cal W}$ around $p=0$ in powers of $k_s$ (since $k_s$ is small when $v$ is small)
to have the following equation,
\begin{align}
\sum_{n=1}^{\infty}B_n v^n=v\sum_{n=0}^{\infty} \left[\frac{{\cal W}^{(n)}}{n!}\right]\left\{\sum_{m=1}^{\infty}B_mv^m\right\}^n\,, \quad \text{with}\quad {\cal W}^{(n)}=\left[\frac{d^n{\cal W}(p)}{dp^n}\right]_{p=0}\,.
\end{align}
Equating the powers of $v$ on both sides of the above equation, one can find the solutions for $B_n$. 
The first few coefficients are,
\begin{equation}\label{solb}
\begin{split}
&B_1 = \left[{\cal W}\right]_{p=0},~~~B_2 = B_1\left[d{\cal W}\over dp\right]_{p=0},~~~B_3 = B_2\left[d{\cal W}\over dp\right]_{p=0}+{1\over 2} B_1^2\left[d^2{\cal W}\over dp^2\right]_{p=0}~.
\end{split}
\end{equation}
\end{itemize}
\end{itemize}
To summarize the discussion on detecting the existence of spurious modes, one can follow the following steps of the algorithm:
\begin{enumerate}
    \item From the relation $k=\gamma\{p-v~{\cal W}(p)\}$, solve for $p(k,v)$.
    \item Calculate $\omega$ from the relation $\omega = \gamma\left\{{\cal W}(p)-v~p\right\}$ using the solution of $p$.
    \item Evaluate $\omega(k,v)$ near $v\to0$ limit.
    \item The $\omega$ solution, with $p=k_s(v) $ (obtained from $\gamma\{p-v~{\cal W}(p)\}=0$) that has a $v$ dependence lower than linear, i.e., $k_s\sim O(v^\alpha)$ with $\alpha<1$, indicates that it is a spurious mode, since it leads to a divergent RHS in \eqref{zeroomega}.
    \item Hence, since $k_s(v)$ does not have a fractional power of $v$, $\omega(k)$ includes a spurious mode unless $[\lim_{v\rightarrow 0} k_s(v)] = 0$. 
\end{enumerate}

\section{Non-spurious boosted modes in terms of local rest frame modes}
\label{nonspurious}
The discussion in Section~\ref{general} is indicative of the fact that there is a set of modes in the boosted frame that are in one-to-one correspondence with the LRF modes and also smoothly reduce to the LRF modes in the limit $v\rightarrow 0$. In this section, we shall present a set of explicit formulae based on the algorithm presented in the previous section for these non-spurious boosted modes in terms of the corresponding LRF modes in an expansion in $k$ or $v$ or both. 
In Section~\ref{general} we have already discussed a unified procedure of computing modes (both for spurious and non-spurious modes) in a boosted frame in terms of LRF modes. But in this section, we are specializing in non-spurious modes, which simplifies the formula. For hydrodynamic modes, it allows us to get an all-order recursive relation for the expansion coefficients, while for non-hydrodynamic modes, we could get the mode equation in a double expansion of $v$ and $k$, to some desirable high orders.
\subsection{Formal set-up: derivation of general formulae}
\label{subsec:formal_general_formulae}

As demonstrated in the previous section, the starting point is the roots or the zeroes of the LRF modes.
We will begin with the assumption that in the LRF, around $\tilde k \to 0$, the modes $\tilde \omega$ can be expressed in an infinite series in $\tilde k$ as follows,
\begin{equation}\label{O}
    \tilde \omega = \sum_{n=0}^\infty a_n \tilde k^n~.
\end{equation}
These modes are nothing but the LRF roots $W^i(p)$ mentioned in Eq.~\eqref{roots} expanded around zero momenta, where the expansion coefficients can be related to the derivative coefficients of the Taylor expansion as $a_n=\left[W_i^{(n)}\over n!\right]$.
The coefficients $a_n$ are constant numbers dependent on the parameters in the underlying microscopic theory. For the hydrodynamic modes, Eq.~\eqref{O} corresponds to $a_0 = 0$, and $a_0 \neq 0$ corresponds to the non-hydrodynamic modes.

Next, we consider the same mode in a Lorentz boosted frame, where its frequency $\omega$ and wavenumber $k$ are related to their LRF counterparts $\tilde \omega$ and $\tilde k$ following \eqref{boost1}. For convenience in the following calculations, we will explicitly express them as,
\begin{align}
    \frac{\omega}{\gamma} &= \tilde \omega - v \tilde k~,  &&\frac{k}{\gamma} = \tilde k - v \,\tilde \omega   \label{B1}~, \\
    \text{and,~~~~~~~~} \frac{\tilde \omega}{\gamma} &= \omega + v k~,  &&\frac{\tilde k}{\gamma} = k + v \,\omega \label{B2}~.
\end{align}
Substituting \eqref{O} in \eqref{B1} and \eqref{B2} we get,
\begin{align}
    \frac{\omega}{\gamma} &= - v \tilde k+\sum_{n=0}^{\infty} a_n \tilde k^n  = a_0 + \frac{ \omega_q}{\gamma} \label{P1}~, \\
    \frac{ k }{\gamma} &= \tilde k - v \sum_{n=0}^\infty a_n \tilde k^n = - v a_0 + \frac{ q }{\gamma} \label{P2}~,
   \end{align}
where we have defined $\omega_q$ and $q$ as,
\begin{align}
    \frac{\omega_q}{\gamma} &= - v \tilde k + \sum_{n=1}^\infty a_n \tilde k^n \label{H1}~, \\
    \frac{ q}{\gamma} &= \tilde k - v \sum_{n=1}^\infty a_n \tilde k^n \label{H2}~.
\end{align}
Note from Eqs.~\eqref{P1} and \eqref{P2}, for the hydrodynamic modes we have $\omega=\omega_q$ and $ k=q$ as the leading coefficient $a_0$ vanishes. Therefore, solving for $ \omega_q$ and $ q$ is sufficient to derive the expansion coefficients for the hydrodynamic modes. Additionally, since we are working with a non-spurious mode here, we do not need to worry about the possibility of multiple solutions in \eqref{P1} and \eqref{P2} as well as in \eqref{H1} and \eqref{H2}.

Here, we proceed to obtain both the hydrodynamic and non-hydrodynamic modes in the boosted frame from the LRF modes using the following steps. The idea here is to first extract $\tilde{k}$ from Eq.~\eqref{H2} in terms of $q$ and substitute it back into Eq.~\eqref{H1} such that $\omega_q$ is expressed in terms of $q$. The right-hand side of \eqref{H2} is a polynomial in $\tilde{k}$ that could have a solution in a power series of $\frac{q}{\gamma}$. Also, from \eqref{H2}, we notice that $q=0$ at $\tilde k=0$. Therefore, we can invert Eq.~\eqref{H2} to write $\tilde k$
in an expansion of $\frac{q}{\gamma}$ in the following power series form,
\begin{equation}\label{B}
    \tilde k = \sum_{n=1}^\infty b_n \left( \frac{ q}{\gamma} \right)^n~,
\end{equation}
where $b_n$ are the unknown coefficients that depend on the boost velocity $v$. The LRF coefficients $a_n$ will be determined from Eq.~\eqref{H2} itself. 

Equation~\eqref{B} connects the LRF momenta $\tilde{k}$ to the boosted momenta $q$ (excluding the dependence from the non-hydrodynamic part of $\tilde{\omega}$ in $k$) and traces back from Eq.~\eqref{ansatz01} of Section \ref{general}, with the coefficients $P_n(v)$ renamed as $b_n$ in this section. Putting Eq.~\eqref{B} into Eq.~\eqref{H2}, we find the following recursive relation for $q$
\begin{align}{\label{BBB}}
\frac{q}{\gamma}=\sum_{m=1}^{\infty}b_m\left(\frac{q}{\gamma}\right)^m-v\sum_{n=1}^{\infty}a_n\left\{\sum_{m=1}^{\infty}b_m\left(\frac{q}{\gamma}\right)^m\right\}^n~.
\end{align}
The coefficients $b_n$ can now be conveniently calculated in terms of the LRF coefficients $a_n$ by comparing the powers of $\frac{q}{\gamma}$ on both sides of Eq.~\eqref{BBB}. One can readily identify Eq.~\eqref{BBB} as the zero momentum limit ($k_s\rightarrow 0$) of Eq.~\eqref{ansatz1} in Section~\ref{general}. The procedure for extracting $b_n$ coefficients from Eq.~\eqref{BBB} has already been explained in Section~\ref{general}, see the discussion around Eq.~\eqref{ansatz1} for extracting the values of $P_n(v)$.

The next task is to evaluate the hydrodynamic modes in the boosted frame as given in Eq.~\eqref{H1}. 
With the calculated values of $b_n$, we use Eq.~\eqref{B} to replace $\tilde{k}$
in Eq.~\eqref{H1} to express $\omega_q$ in terms of $q$
\begin{align}
\label{hydro1}
    \frac{\omega_q}{\gamma}=-v\sum_{m=1}^{\infty}b_m\left(\frac{q}{\gamma}\right)^m+\sum_{n=1}^{\infty}a_n\left\{\sum_{m=1}^{\infty} b_m\left(\frac{q}{\gamma}\right)^m\right\}^n~.
\end{align}
Equation~\eqref{hydro1} states that $\frac{\omega_q}{\gamma}$ must be expressed in a basis of $\frac{q}{\gamma}$, which can be represented in the following infinite series,
\begin{equation}\label{W}
    \frac{\omega_q}{\gamma} = \sum_{n=1}^\infty \tilde a_n \left(\frac{ q}{\gamma}\right)^n~.
\end{equation}
Equation~\eqref{W} is the aimed hydrodynamic mode in the boosted frame, with $\tilde{a}_n$ as the associated dispersion coefficients. Comparing the power of $\frac{q}{\gamma}$ from Eq.~\eqref{hydro1} and \eqref{W}, the coefficients of the boosted hydrodynamic mode $\tilde a_n$ can be determined in terms of $b_n$, $a_n$, and $v$. It is important to note that, with the knowledge of $b_n$ in terms of $a_n$, which we already have, the coefficients $\tilde{a}_n$ can solely be evaluated from the LRF coefficients $a_n$ and boosted velocity $v$. This will be discussed in more detail in the following subsections.

Next, we discuss the calculation of the non-hydrodynamic modes in the boosted frame from the LRF modes. From \eqref{P2}, we have $$\frac{ q}{\gamma}=\frac{ k}{\gamma} +  v a_0\,.$$ 
Using above equation in \eqref{W} and then substituting it in Eq.~\eqref{P1}, we finally obtain,
\begin{align}
       \frac{ \omega}{\gamma} &= a_0 + \sum_{n=1}^\infty \tilde a_n \left( \frac{ k}{\gamma} + v a_0 \right)^n~.
       \label{F11}
\end{align}
A bit of rearrangement of Eq.~\eqref{F11} can reveal that it is exactly the relation in Eq.~\eqref{pertomega} of Section~\ref{general}, just rewritten in a convenient form.
Next, in order to evaluate the boosted non-hydrodynamic modes, the right-hand side of Eq.~\eqref{F11} is expanded over $(a_0v)$ as
\begin{align}
    \frac{\omega}{\gamma}=&\left\{a_0+\tilde{a}_1(va_0)+\tilde{a}_2(va_0)^2+\tilde{a}_3(  va_0)^3+\cdots\infty\right\} \left(\frac{{k}}{\gamma}\right)^0\nonumber\\
+&\left\{\tilde{a}_1+2\tilde{a}_2(va_0)+3\tilde{a}_3(va_0)^2+4\tilde{a}_4(v a_0)^3+\cdots\infty\right\} \left(\frac{{k}}{\gamma}\right)^1\nonumber\\
+&\left\{\tilde{a_2}+3\tilde{a}_3(  va_0)+6\tilde{a}_4(va_0)^2+10\tilde{a}_5(va_0)^3+\cdots\infty\right\} \left(\frac{{k}}{\gamma}\right)^2\nonumber\\
+&\cdots+
\left\{\sum_{m=r}^{\infty}\tilde{a}_m ~ ^m{\text{C}}_r (  va_0)^{m-r}\right\}\left(\frac{{k}}{\gamma}\right)^r+\cdots\infty~.
\label{disp111}
\end{align}
Thus, Eq.~\eqref{disp111} is the non-hydrodynamic mode (more detailed derivation with further organised form will be given in the specific subsection) in the boosted frame, which we write in the consolidated form as,
\begin{equation}
   \frac{\omega}{\gamma}   = \sum_{n=0}^\infty a_n^* \left( \frac{ k}{\gamma}  \right)^n~,
        \label{F}
\end{equation}
with $a_n^*$ as the associated dispersion coefficients. They can again be systematically evaluated by comparing the powers of $\frac{k}{\gamma}$ in Eq.~\eqref{F} and \eqref{disp111}, which turn out to be functions of $\tilde{a}_n$ that, in turn, depend on $v$ and $a_n$.

The connection between Eq.~\eqref{disp111} and Eq.~\eqref{pertomega} (in Section~\ref{general}) can be easily established by comparing the $k$ powers in the $\omega$ expression with this small exercise.

Let us compare the first $\left(\frac{k}{\gamma}\right)^0$ term in $\omega$. Using the explicit values of $\tilde{a}_n$ (documented in the following subsections) in \eqref{disp111},
as well as using \eqref{solb} in Eq.~\eqref{pertomega} in Section~\ref{general}, and finally doing the necessary small $v$ expansion in both the cases upto ${\cal{O}}(v^2)$, we find the leading term in $\omega$ as,
\begin{align}
    \omega=\left\{a_0+(a_0a_1) v + \left(a_0a_1^2+a_2a_0^2-\frac{1}{2}a_0\right)v^2+{\cal{O}}(v^3)\right\}\left(\frac{k}{\gamma}\right)^0+{\cal{O}}\left(\frac{k}{\gamma}\right)^1  ~.
    \label{check}
\end{align}
Conversely, if we substitute $k=0$ in Eq.~\eqref{pertomega} and compare with Eq.~\eqref{check},
then, we obtain an expression for $\frac{k_s(v)}{v\gamma}$ in an expansion of $v$ which can be written as,
\begin{align}
    \left[\frac{k_s(v)}{v\gamma}\right]_{k=0}=a_0+(a_0a_1) v + \left(a_0a_1^2+a_2a_0^2-\frac{1}{2}a_0\right)v^2+{\cal{O}}(v^3)~.
    \label{ks3}
\end{align}
From here, we could derive an expression for $k_s(v)$ in powers of $v$. Following the ansatz given in Eq.~\eqref{ksexpansion} in Section~\ref{general}, which we repeat here for convenience,
$$k_s(v)=\sum_{n=1}B_n \, v^n\,,$$ 
computing the $B_n$ coefficients will do the needful. Substituting this ansatz in Eq.~\eqref{ks3} and comparing the powers of $v$ from both sides, we obtain the first three $B_n$ coefficients
\begin{align}
B_1=a_0~~,~~~~B_2=a_0a_1~~,~~~~B_3=a_0a_1^2+a_2a_0^2~.
\label{Bcoeff}
\end{align}
One can easily notice that these are the exact same coefficients given in Eq.~\eqref{solb} of Section~\ref{general}, if the LRF roots are adopted from Eq.~\eqref{O}.

Note that, for the LRF hydrodynamic modes $(a_0=0)$, Eq.~\eqref{F} reduces to Eq.~\eqref{W}, and we have $a_n^*=\tilde a_n$. For the non-hydrodynamic modes, this doesn't happen due to the non-vanishing $a_0 \neq 0$.
This also shows us that under a Lorentz boost, a non-spurious hydrodynamic mode can never transform into a non-hydrodynamic mode.

In the rest of this section, we will derive all of these expansion coefficients $b_n$, $\tilde a_n$, and $a^*_n$ in terms of $a_n$ and $v$. To avoid cluttering in the expressions, we will use the following notations for the rest of this section,
\begin{equation*}
    x \equiv \frac{q}{\gamma},\qquad y \equiv \frac{k}{\gamma}\,,
\end{equation*}
hence, we express \eqref{B} and \eqref{F} as
\begin{equation*}
    \tilde k =\sum_{n=1}^{\infty}b_n x^n = x\sum_{n=0}^\infty b_{ n+1} x^n,~~~~~~~ \frac{\tilde \omega}{\gamma} = \sum_{n=0}^\infty a_n^* y^n~.
\end{equation*}
\subsection{Calculating coefficients \texorpdfstring{$b_n$}{} as functions of \texorpdfstring{$a_n$}{} and \texorpdfstring{$v$}{}}

We know from \eqref{B} that,
\begin{equation}\label{BP}
    \tilde k = x\sum_{n=0}^\infty b_{ n+1} x^n ~~.
\end{equation}
Using this, we can rewrite \eqref{H2} as ,
\begin{equation}\label{S1}
    \begin{split}
        x &= \tilde k -v \sum_{n=1}^\infty a_n \tilde k^n = \tilde k -  v \tilde k \sum_{m=0}^\infty a_{  m+1} \tilde k^m ~,\\
        &= x \sum_{n=0}^\infty b_{  n+1} x^n -  v \left( x \sum_{n=0}^\infty b_{  n+1} x^n \right) \left\{ \sum_{m=0}^\infty a_{  m+1}x^m  \left(  \sum_{l=0}^\infty b_{l+1} x^l \right)^m\right\} ~.\\
        \Rightarrow -1 &+  \sum_{n=0}^\infty b_{  n+1} x^n -  v \left[ \left(  \sum_{n=0}^\infty b_{  n+1} x^n \right) \left\{ \sum_{m=0}^\infty a_{  m+1}x^m  \left(  \sum_{l=0}^\infty b_{l+1} x^l \right)^m\right\}\right]  = 0 ~,\\
        \Rightarrow -1 &+ \sum_{n=0}^\infty b_{  n+1} x^n - v Q_1  = 0~,
    \end{split}
\end{equation}
where we have defined $Q_1$ in the following way for convenience
\begin{equation}
    \begin{split}
        Q_1 &= Q_2 Q_3 ~,~~~~ Q_2 = \left\{  \sum_{n=0}^\infty b_{  n+1} x^n \right\}~, \\
        Q_3 &= \left\{ \sum_{m=0}^\infty a_{  m+1}x^m  Q_4\right\}~,~~~~Q_4 = \left\{\sum_{l=0}^\infty b_{l+1} x^l \right\}^m~.
    \end{split}
\end{equation}
For calculating $b_{n+1}$, we would ultimately resort to power counting of the terms in \eqref{S1}. We'll start this from $x^0$ onwards. In the coefficient of $x^r$ in \eqref{S1}, for $r\neq 0$, contributions come from the second and third terms. The second term gives a $b_{r+1}$, while the third term provides a more complicated expression. We can think of this in terms of collecting a total of $r$ no. of $x$ powers from the terms in $Q_1$. One extreme case occurs when all the $r$ no. of powers are collected from $Q_2$. Then we get the leading term as $b_{r+1}$ and only $a_1$ contributes from $Q_3$. Another extreme case arises when $Q_2$ contributes $0$ powers and $Q_3$ contributes all the $r$ no. of powers. Then we observe that any $b_{r+1}$ contributions from $Q_4$ must be with $n<r$ as some non-zero power contribution would always come from the $a_{m+1}$. 

Therefore, in the expression of $b_{r+1}$ that we want to calculate from the coefficient of $x^r$, there are no leading order (i.e., $b_{r+1}$) contributions from the $Q_3$ part. For the time being, we will refrain from getting into the details of the exact form of $Q_3$ and represent it schematically in the $x$-expansion as 
\begin{equation}\label{eqcj}
    \sum_{m=0}^\infty a_{  m+1} x^m \left(  \sum_{l=0}^\infty b_{l+1} x^l \right)^m = \sum_{j=0}^\infty c_j x^j ~,
\end{equation}
with $c_0 = a_1$. These $c_j$ can be read off by expanding the LHS and RHS and comparing the powers of $x^r$. Then, using the Cauchy product rule:
$$\left(\sum_{n=0}^\infty {\cal{A}}_n \right) \left(\sum_{l=0}^\infty {\cal{B}}_l \right) = \sum_{n=0}^\infty \left\{ \sum_{l=0}^n {\cal{A}}_l {\cal{B}}_{n-l} \right\}\,,$$ 
we can write $Q_1$ as
\begin{equation}\label{Q1}
    Q_1 = \left\{ \sum_{n=0}^\infty b_{n+1}x^n  \right\} \left\{ \sum_{m=0}^\infty c_m x^m \right\} = \sum_{j=0}^\infty d_j x^j,\quad \text{where} \quad d_j = \sum_{n=0}^j b_{n+1} c_{j-n}~.
\end{equation}
Substituting this into \eqref{S1}, we get
\begin{equation}\label{S2}
    \sum_{n=0}^\infty -\delta_{n0} + \sum_{n=0}^\infty b_{  n+1}x^n - v \sum_{n=0}^\infty d_n x^n = 0~.
\end{equation}
Now, from the coefficient of $x^r$ in \eqref{S2}, we obtain
\begin{equation}
    \begin{split}
        &- \delta_{r0} + b_{r+1} -  v d_r = 0 ~, ~~~\Rightarrow - \delta_{r0} + b_{r+1} - v \sum_{n=0}^r b_{n+1} c_{r-n} = 0 ~,\\
        &\Rightarrow -\delta_{r0} + (1- v~c_0) b_{r+1} -  v \sum_{n=0}^{r-1}b_{n+1} c_{r-n} = 0~.
    \end{split}
\end{equation}
This is a recursion relation for $b_{r+1}$. So, if we know all the previous $b_n$ up to $n=r$ in terms of $a_n$, then we can calculate $b_{r+1}$ in terms of $a_n$ and $v$. Using the fact that $c_0=a_1$, we can solve for the $b_r$ as,
\begin{align}
b_{r+1} = \frac{1}{(1- v~c_0)} \left[ \delta_{r0} + v \sum_{n=0}^{r-1} b_{n+1} c_{r-n} \right]~, ~~~~~r\geq 0,
\end{align}
which gives
\begin{align}
    b_1=\frac{1}{1- va_1}~,~~~~~~b_{r+1} &= \frac{ v}{(1- v~a_1)} \sum_{n=0}^{r-1} b_{n+1} c_{r-n}~,~~~~~~~(k\geq 1)~.
\end{align}
For convenience, we are listing the first few values of $b_n$ in terms of the LRF coefficients $a_n$
\begin{align}
 b_1=&\frac{1}{(1- va_1)}~,~~~~b_2=\frac{va_2}{(1- va_1)^3}~,~~~~b_3=\frac{v \left[a_3 \left(1-a_1 v\right)+2 a_2^2 v\right]}{\left(1-a_1 v\right){}^5}~,\nonumber\\
 b_4=&\frac{v \left[a_4 \left(1-a_1 v\right)^2+5 a_2^3 v^2+5 a_3 a_2 v \left(1-a_1 v\right)\right]}{\left(1-a_1 v\right)^7}~.
\end{align}
One can notice that the first three $b_n$ values are identical to the $P_n(v)$ values (Eq.~\eqref{solp1}).
\subsection{Calculation of hydrodynamic coefficients \texorpdfstring{$\tilde a_n$}{}}

We now rewrite \eqref{W} as
\begin{equation}\label{WP}
    \frac{ \omega_q}{\gamma} = \sum_{n=1}^\infty \tilde a_n x^n = x \sum_{n=0}^\infty  \tilde{a}_{  n+1}x^n~,
\end{equation}
and using \eqref{BP}, we can rewrite \eqref{H1} as
\begin{equation}\label{S3}
    \begin{split}
        \frac{\omega_q}{\gamma} &= - v \tilde k+ \sum_{n=1}^\infty a_n \tilde k^n = - v \tilde k + \tilde k \sum_{n=0}^\infty a_{  n+1}\tilde k^n ~,\\
        \text{which follows,}\\ \Rightarrow x \sum_{n=0}^\infty \tilde a_{  n+1} x^n &= - v x \sum_{n=0}^\infty b_{  n+1}x^n + \left( x \sum_{n=0}^\infty b_{  n+1}x^n \right) \left\{ \sum_{m=0}^\infty a_{  m+1}x^m \left( \sum_{l=0}^\infty b_{l+1}x^l \right)^m \right\}~, \\
        \Rightarrow \sum_{n=0}^\infty \tilde a_{  n+1}x^n &= - v \sum_{n=0}^\infty b_{  n+1} x^n + \left[ \left( \sum_{n=0}^\infty b_{  n+1}x^n \right) \left\{ \sum_{m=0}^\infty a_{  m+1}x^m \left( \sum_{l=0}^\infty b_{l+1}x^l \right)^m \right\} \right] ~,\\
        \Rightarrow \sum_{n=0}^\infty \tilde a_{  n+1}x^n &= -v \sum_{n=0}^\infty b_{  n+1} x^n + Q_1 ~.
    \end{split}
\end{equation}
Substituting \eqref{Q1} in the last line of \eqref{S3} we get
\begin{equation}\label{S4}
    \Rightarrow \sum_{n=0}^\infty \tilde a_{  n+1}x^n = - v \sum_{n=0}^\infty b_{  n+1} x^n + \sum_{n=0}^\infty d_n x^n~.
\end{equation}
The coefficient of the $x^r$ term in \eqref{S4} can then be written as
\begin{equation}
    \begin{split}
        \tilde a_{r+1} &= - v~ b_{r+1} + d_r = - v ~ b_{r+1} + \sum_{n=0}^r b_{  n+1} c_{r-n} 
        = (- v+ c_0) b_{r+1} + \sum_{n=0}^{r-1} b_{  n+1} c_{r-n}\,, \\
        \Rightarrow \tilde a_{r+1} &= (- v+a_1) b_{r+1} + \sum_{n=0}^{r-1} b_{  n+1} c_{r-n}\,.
    \end{split}
\end{equation}
Using previously derived expressions of $b_{r+1}$, we obtain
\begin{equation}
    \tilde a_{r+1} = \frac{- v+a_1}{1- v~a_1} \delta_{r0} + \left\{ \frac{ v(- v+a_1)}{1- v~a_1} - 1\right\} \sum_{m=0}^{r-1} b_{  m+1} c_{r-m}~.
\end{equation}
After some mathematical simplifications given below,
\begin{equation}
    \begin{split}
        &\frac{-v+a_1}{1- v~a_1} = - v + (1- v^2) \frac{a_1}{1- v~a_1} = - v + \frac{1}{\gamma^2} \frac{a_1}{1- v~a_1}~, \\
        &1 + \frac{ v(-v+a_1)}{1- v~a_1} = \frac{1- v^2}{1- v~a_1} = \frac{1}{\gamma^2} \frac{1}{1- v~a_1}~,
    \end{split}
\end{equation}
we express $\tilde a_{r+1}$ as, 
\begin{align}\label{AT}
    \tilde a_{r+1} &= - v \delta_{r0} + \frac{1}{\gamma^2 (1- v~a_1)} \left[ a_1 \delta_{r0} + \sum_{m=0}^{r-1} b_{m+1} c_{r-m} \right]~,~~~(r \geq 0)~.
\end{align}
Note that the first term purely contributes to $\tilde a_1$, and the second term purely contributes to $\tilde a_{r\geq 2}$. So we have,
\begin{align}
    \tilde a_1 &= - v + \frac{1}{\gamma^2} \frac{a_1}{(1- v~a_1)}~,\\
    \tilde a_{r+1} &= \frac{1}{\gamma^2} \frac{1}{(1- v~a_1)} \sum_{m=0}^{r-1} b_{m+1} c_{r-m}~,~~~ (r \geq 1).
\end{align}
Thus, we observed that the $\tilde{a}_n$ coefficients can be expressed in terms of $b_n$, which, in turn, can be expressed in terms of $a_n$ as shown in the previous subsection.
We are listing here the first few values of $\tilde{a}_n$ in terms of the LRF coefficients $a_n$
\begin{align}
 \tilde{a}_1=&-v+\frac{a_1}{\gamma ^2 \left(1-a_1 v\right)}~,~~~~\tilde{a}_2=\frac{a_2}{\gamma ^2 \left(1-a_1 v\right){}^3}~,~~~~\tilde{a}_3=\frac{a_3 \left(1-a_1 v\right)+2 a_2^2 v}{\gamma ^2 \left(1-a_1 v\right){}^5}~,\nonumber\\
 \tilde{a}_4=&\frac{a_4 \left(1-a_1 v\right){}^2+5 a_2 v \left\{a_2^2 v+a_3 \left(1-a_1 v\right)\right\}}{\gamma ^2 \left(1-a_1 v\right){}^7}~.
\end{align}

\subsection{Calculation of non-hydrodynamic coefficients \texorpdfstring{$a_n^*$}{} }

We now move on and rewrite \eqref{P1} and \eqref{P2} as 
\begin{equation}
    x = y + v~a_0\,, \qquad \frac{ \omega}{\gamma} = a_0 + \frac{ \omega_q}{\gamma}~.
\end{equation}
Substituting in Eqs.~\eqref{W} and \eqref{F} we get
\begin{equation}
    \frac{\omega}{\gamma}=\sum_{n=0}^\infty a_n^* y^n = a_0 + \sum_{n=1}^\infty \tilde a_n (y+v~a_0)^n \equiv f(y)~.
\end{equation}
Using the Taylor expansion, we can express it as
\begin{equation}
    a_m^* = \frac{1}{m!} \left[ \left( \frac{d}{dy} \right)^m \left\{ a_0 + \sum_{n=1}^\infty \tilde a_n (y+ v~a_0)^n \right\} \right]_{y=0}~.
\end{equation}
Furthermore, we can obtain the following relations
\begin{equation*}
    \begin{split}
        \frac{d^m}{dy^m}a_0 = a_0 \delta_{m0} ~~~&\to~~~ \frac{1}{m!}\frac{d^m}{dy^m}a_0 = a_0 \delta_{m0}~,\\
        \left[ \left( \frac{d}{dy} \right)^m \sum_{n=1}^\infty \tilde a_n (y+ v~a_0)^n  \right]_{y=0} &= \sum_{n=m}^\infty \frac{n!}{(n-m)!} \tilde a_{n+1} (v~a_0)^{n+1-m}~.
    \end{split}
\end{equation*}
Therefore, $a_m^*$ is written as
\begin{equation}
    \begin{split}
        a_0^* &= a_0 + \sum_{n=1}^{\infty} \tilde a_n ( v~a_0)^{n}~,\\
        a_m^* &= \sum_{n=m}^\infty {n\choose m} \tilde a_{n} ( v~a_0)^{n-m}~, \quad m\geq1~.
    \end{split}
\end{equation}
From the above set of equations, we observe that for $a_0=0$, $a_0^*=0$, and $a_n^*=\tilde{a}_n$ for $n\geq1$, as hydrodynamic modes are expected to be recovered.
After some simplifications in the boosted frame, we have the following dispersion relation
\begin{align}
\left\{\frac{{\omega}}{\gamma}+ v\frac{{k}}{\gamma}\right\}=
&\left[\frac{a_0}{\gamma^2}(  va_0)^0+\tilde{\tilde{a_1}}(  va_0)^1+\tilde{a}_2(  va_0)^2+\tilde{a}_3(  va_0)^3+\cdots\infty\right] \left(\frac{{k}}{\gamma}\right)^0\nonumber\\
+&\left[\tilde{\tilde{a_1}}(  va_0)^0+2\tilde{a}_2(  va_0)^1+3\tilde{a}_3(  va_0)^2+4\tilde{a}_4(  va_0)^3+\cdots\infty\right] \left(\frac{{k}}{\gamma}\right)^1\nonumber\\
+&\left[\tilde{a_2}(  va_0)^0+3\tilde{a}_3(  va_0)^1+6\tilde{a}_4(  va_0)^2+10\tilde{a}_5(  va_0)^3+\cdots\infty\right] \left(\frac{{k}}{\gamma}\right)^2\nonumber\\
+&\cdots+
\left[\sum_{m=r}^{\infty}\tilde{a}_m~{}^m{\text{C}}_r (  va_0)^{m-r}\right]\left(\frac{{k}}{\gamma}\right)^r+\cdots\infty~,
\label{disp1}
\end{align}
with $\tilde{\tilde{a_1}}=\tilde{a}_1+ v=\frac{1}{\gamma^2}\frac{a_1}{1- va_1}$.
Here we observe that each coefficient of $(  va_0)^n$ in each $\left(\frac{{k}}{\gamma}\right)^m$ basis
is proportional to $\frac{1}{\gamma^2}$. So alternatively, in a consolidated form, Eq.~\eqref{disp1} can be written as
\begin{align}
    \gamma\left({\omega}+v{k}\right)=\sum_{n=0}^{\infty}\alpha_n^* \left\{\frac{{k}}{\gamma}\right\}^n \,,\qquad \alpha_n^*=\gamma^2 \sum_{m=n}^{\infty}\alpha_m~{}^{m}{\text{C}}_{n}~(  va_0)^{m-n}~,
    \label{disp2}
\end{align}
where,
\begin{align}
\alpha_0=\frac{a_0}{\gamma^2}~;~~~~\alpha_1=\tilde{\tilde{a}}_1~;~~~~~\alpha_n=\tilde{a}_n~~{\text{for}}~n\geq 2~.
\end{align}
Equation~\eqref{disp2} is the main result of this subsection.

There are several important points to be noted here. Firstly, we observe that in a boosted frame, the dispersion relations can be fully understood from the same information in the local rest frame itself. Using \eqref{disp2}, the expansion coefficients of the non-spurious modes can be derived using only the $a_n$ in \eqref{O}, and the boost parameter $v$. This allows us to bypass the traditional method of deriving these coefficients by calculating the dispersion polynomial in the boosted frame and solving for its roots and their expansions anew. Next, we observe that for hydrodynamic modes in the boosted frame, the $\tilde{a}_n$ coefficients comprise a combination of $v$ and only a finite number of $a_n$. For the non-hydrodynamic modes, however, every $a_n^*$ is itself an infinite series in $(v a_0)$. The coefficients of this nested series are made up of $v$ and the $a_n$ that includes $n$ ranging from $0$ to $\infty$, i.e., the knowledge of the LRF modes to its entirety.

Finally, we would like to mention an interesting feature observed in the previous analysis. We find that in the boosted frame, $\omega$ is obtained not merely in a power series of $k$, but rather in a series of $(k /\gamma)$. At the limit of $v \to 1, \gamma^{-1} \to 0$, therefore, many of the sub-leading terms in this expansion become insignificant at an ultra-high boost. This phenomenon of \emph{$\gamma$-suppression} leads to profound implications for the stability and causality-related properties of the fluid in a Lorentz boosted frame~\footnote{A similar interplay of stability and causality criteria at an ultra-high boost was studied in Ref.~\cite{Roy:2023apk}. The calculation presented here hints at the reason why the results derived there at ultra-high boosts will remain valid even at $k\neq 0$.}. This connection will be explored in greater detail in an upcoming work~\cite{Roy:2026tbr}.

\subsection{A simple example: An equation of the Maxwell-Cattaneo type}

In this subsection, we will apply the previous discussion to the Maxwell-Cattaneo type equation~\cite{Romatschke:2009im}. This is the same equation that appears in the shear channel of the Muller-Israel-Stewart (MIS) theory. The solution to this equation contains two modes (one hydrodynamic and one non-hydrodynamic), in both of which we will apply our analysis of the non-spurious modes and show that the expansion coefficients of their dispersion relations in the boosted frame are, indeed, determined by those in the LRF. We will further show how the ``$\gamma$-suppression" that we discussed earlier occurs in both the hydrodynamic and the non-hydrodynamic modes.

The Maxwell-Cattaneo equation can be expressed as,
\begin{equation}\label{maxcat}
    \tilde{\omega}^2 + \frac{i}{\tau_\Pi} ~ \tilde{\omega} - \frac{\lambda}{\tau_\Pi}\tilde{k}^2 = 0~,
\end{equation}
where $\lambda$ and $\tau$ are the constant parameters arising from the underlying theory.
The two solutions to this equation are given by
\begin{equation}
    \tilde{\omega} = \frac{-\frac{i}{\tau_\Pi}\pm \sqrt{ -\frac{1}{\tau_\Pi^2} +\frac{4\lambda}{\tau_\Pi}\tilde{k}^2 }}{2}~.
\end{equation}
    The solution for the non-hydrodynamic mode can be expressed as a series expansion around $\tilde{k}=0$
\begin{equation}\label{maxcatnh}
    \tilde{\omega} = -\frac{i}{\tau_\Pi } + i \lambda~\tilde{k}^2 + i \lambda ^2\tau_\Pi~\tilde{k}^4 + 2 i \lambda ^3  \tau_\Pi ^2~ \tilde{k}^6 + O\left(\tilde{k}^7\right)~~.
\end{equation}
Similarly, solutions for the hydrodynamic mode can be written as
\begin{equation}\label{maxcathydro}
    \tilde{\omega} = -i \lambda~\tilde{k}^2 - i \lambda ^2  \tau_\Pi~\tilde{k}^4  - 2 i \lambda ^3  \tau_\Pi ^2~\tilde{k}^6 + O\left(\tilde{k}^7\right)~~.
\end{equation}
First, we will extract the boosted modes corresponding to \eqref{maxcatnh} and \eqref{maxcathydro} in the traditional way, i.e., by boosting the polynomial. For this, we go to a boosted frame following \eqref{boost1}, under which the dispersion polynomial \eqref{maxcat} becomes,
\begin{equation}\label{tradboost}
    \omega^2 -\frac{i }{\gamma(\lambda  v^2-\tau_\Pi) }~\omega +\frac{2 v (\lambda -\tau_\Pi )}{\lambda  v^2-\tau_\Pi }~\omega k - \frac{i v }{\gamma(\lambda  v^2-\tau_\Pi )}~ k + \frac{\lambda -\tau_\Pi  v^2}{\lambda  v^2-\tau_\Pi }~k^2 = 0~.
\end{equation}
Solving Eq.~\eqref{tradboost}, we can obtain the boosted modes.
The dispersion relation for the non-hydrodynamic mode becomes
\begin{equation}\label{tradnh}
    \begin{split}
        \omega =~& 
        \frac{i}{\gamma  \left(\lambda  v^2-\tau_\Pi \right)}
        + \frac{ v \left(\tau_\Pi +\lambda  \left(v^2-2\right)\right)}{\lambda  v^2-\tau_\Pi }~k
        + \frac{i \lambda  }{\gamma ^3}~k^2
        + \frac{2  \left(\lambda ^2 v\right)}{\gamma ^4} ~k^3\\
        &- \frac{i \lambda ^2  \left(5 \lambda  v^2-\tau_\Pi \right)}{\gamma ^5}~k^4
        + O\left(k^5\right)~,
    \end{split}
\end{equation}
and that for the hydrodynamic mode becomes
\begin{equation}\label{tradh}
    \begin{split}
        \omega =~& - v~k
        - \frac{i \lambda  }{\gamma ^3}~k^2
        - \frac{2 \lambda ^2  v}{\gamma ^4}~k^3
        + \frac{i \lambda ^2  \left(5 \lambda  v^2-\tau_\Pi \right)}{\gamma ^5}~k^4+ O\left(k^5\right)~.
    \end{split}
\end{equation}
Next, we will reproduce the boosted modes using the method described in the previous section. For that, from the dispersion coefficients of the LRF dispersion modes given in Eq.~\eqref{maxcatnh} and \eqref{maxcathydro}, we need to systematically calculate the coefficients $\tilde{a}_n$ (for the hydrodynamic mode) and $a^{*}_n$ (for the non-hydrodynamic mode). Here, we are quoting the final results for the boosted modes.

The boosted non-hydrodynamic mode gives, 
\begin{align}\label{newnh}
    \gamma\left\{\omega+vk\right\}=&\frac{i}{\left(v^2\lambda-\tau_{\pi}\right)}\left(\frac{k}{\gamma}\right)^0
    -\frac{2v\lambda}{\left(v^2\lambda-\tau_{\pi}\right)}\left(\frac{k}{\gamma}\right)^1
    +~i\lambda\left(\frac{k}{\gamma}\right)^2+~2v\lambda^2\left(\frac{k}{\gamma}\right)^3\nonumber\\
    &-~i\lambda^2\left(5v^2\lambda-\tau_{\pi}\right)\left(\frac{k}{\gamma}\right)^4+~{\cal{O}}\left(\frac{k}{\gamma}\right)^5~,
\end{align}
whereas the boosted hydrodynamic mode gives
\begin{align}\label{newh}
  \gamma\{\omega+vk\}=-i\lambda\left(\frac{k}{\gamma}\right)^2-2v\lambda^2  \left(\frac{k}{\gamma}\right)^3+
  i\left\{5v^2\lambda^3-\tau_{\pi}\lambda^2\right\}\left(\frac{k}{\gamma}\right)^4+{\cal{O}}\left(\frac{k}{\gamma}\right)^5~.
\end{align}
A simple algebraic rearrangement and an appropriate expansion of each coefficient of $(k/\gamma)$ in terms of $v$ will show that the boosted modes \eqref{newnh} and \eqref{newh} derived using the current method are identical to those calculated in \eqref{tradnh} and \eqref{tradh}, respectively, using the traditional technique of polynomial boosting. \\
\\
To summarize:
\begin{itemize}
    \item This exercise demonstrates that, in the current work, we are providing a unique technique to obtain the dispersion spectra in the boosted frame, bypassing the traditional method of boosting the entire LRF polynomial and then solving it. In the current approach, the sole information of the LRF modes will suffice to achieve this. Here, we would like to mention that solving the dispersion polynomial in an arbitrary reference frame could be extremely nontrivial~\cite{Gavassino:2021owo}, especially when, unlike this simple test case (Eq.~\eqref{maxcat}), the theory involves a much higher-order of the LRF polynomial. In that respect, the current analysis could be quite effective in evaluating the boosted spectra of a given theory.     
    \item The feature of  ``$\gamma$ suppression" mentioned earlier is clearly depicted in \eqref{newnh} and \eqref{newh} as the boosted modes are obtained in a power series of $\frac{k}{\gamma}$. In near luminal boosts, this feature diminishes all the sub-leading terms in the frequency mode, leading to significant effects on causality and stability, which will be investigated further in our upcoming work~\cite{Roy:2026tbr}. 
\end{itemize}


\section{The existence of the spurious mode and the impact on causality}
\label{spurious}
It is well known that under boost transformations, it is possible to generate new modes in the theory~\cite{Hoult:2023clg,Denicol:2008ha,Mitra:2021ubx}, which might violate causality and/or the stability of the boosted equilibrium, despite the theory having satisfied stability or asymptotic causality criteria in the LRF.
Clearly, for any Lorentz-invariant theory, a simple Lorentz transformation should never lead to such a drastic change. And, in Ref.~\cite{Hoult:2023clg}, the authors have indeed claimed that the mere existence of such extra modes in the boosted frame is actually connected to the causality violation in the underlying fluid theory. 

In this section, we shall explore the same connection between the violation of causality and spurious modes from a different perspective, namely the many-to-one mapping from the complex $p$ (momentum in LRF) plane to the complex $k$ (momentum in the boosted frame) plane via the first equation of~\eqref{solpara2}.
\subsection{Zero-boost limit of the spurious modes}
Let us first show that whenever the boosted polynomial $P_v(\omega,k,v)$ has more zeroes than the LRF (i.e., $P(\omega,k)$), the extra zeroes of $P_v$ must diverge in the $v\rightarrow0$ limit. In the following steps, we will show in detail.

The most general form of $P(\tilde \omega,\tilde k)$ - the dispersion polynomial in LRF is the following~\footnote{We request the reader not to confuse the coefficients $a_m$ and $b_n^{(m)}$ used in this section with the similar notation used in Section~\ref{nonspurious}. The notations used for dispersion coefficients in each section limit their application to that particular section unless mentioned generically in Subsection~\ref{notcon}.}
\begin{equation}\label{mostgen}
P(\tilde \omega,\tilde k)= \sum_{m=m_0}^M a_m~\tilde\omega^m \left(\sum_{n=0}^{N_m} b^{(m)}_n \tilde k^n\right)~.
\end{equation}
For any fixed complex $k$, this polynomial will have $M$ zeroes of the form $\tilde\omega = W^i(\tilde k), ~~i\in\{1,2,\cdots,M\}$. These $M$ zeroes will determine the $M$ modes of the linearized perturbation in the LRF.

Now, in the  boosted frame, the structure of the polynomial would be,
\begin{equation}\label{mostgenboost}
P_v(\omega, k,v)=\sum_{m=m_0}^M a_m~\gamma^m\left(\omega + v ~k\right)^m \left\{\sum_{n=0}^{N_m} b^{(m)}_n\gamma^n\left(k + v~\omega\right)^n\right\}~.
\end{equation}
Let $M_v$ denote the highest power of $\omega$ in $P_v(\omega,k,v)$ and therefore its number of zeroes. Inspecting equation \eqref{mostgenboost} we could easily see that,
$$M_v = \text{ maximum of $(m+n)$ such that $a_m b_n^{(m)}\neq 0$}~\geq~ M~.$$ 
It is clear that whenever $M_v>M$, the theory will have spurious modes (i.e., modes that appear only when $v\neq0$)~\footnote{This condition has also been stated in Eq.~(6) of Ref.~\cite{Hoult:2023clg}.} and the number of such spurious modes will be $N_{sp}\equiv M_v -M$.
Since at $v\rightarrow 0$, $P_v(\omega,k,v)$ and $P(\tilde{\omega},\tilde{k})$ are identical, the polynomial $P_v(\omega,k,v)$ could be expanded as a finite power series in $v$ as
\begin{equation}
    P_v(\omega,k,v) = P(\gamma~\omega,\gamma~k) +\sum_{r=1}^{r_{\rm max}} v^r Q^{(r)}(\gamma~\omega,\gamma~k)~.
\label{Q}
\end{equation}
Let $\omega =W_v^i(v,k)$ is a zero of $P_v(\omega,k,v)$ such that, $\lim_{v\rightarrow0} W_v^i(v,k)=\text{finite}$, i.e., we are considering a non-spurious mode. Then, using Eq.~\eqref{Q}, the zeroes of the boosted polynomial can be expressed in the following manner
\begin{equation}
0=P_v(W_v^i(v,k),k)= P(\gamma W_v^i(v,k),\gamma k) +\sum_{r=1}^{r_{\rm max}} v^r Q^{(r)}(\gamma W_v^i(v,k),\gamma k)~.
\label{show01}
\end{equation}
Taking the limit $v\rightarrow 0$, it readily follows that,
\begin{equation}
\lim_{v\rightarrow0}\sum_{r=1}^{r_{\rm max}} v^r Q^{(r)}\left(\gamma W_v^i(v,k),\gamma~k\right) = 0~.
     \label{Q11}
\end{equation}
Further, taking successive  $v$ derivatives of equation \eqref{show01} and evaluating them at the limit of $v\rightarrow0$, we could show that all $v$ derivatives of $W_v(v,k)$ are also finite at $v=0$ and uniquely fixed by the zero boost limit of $W_v(v,k)$. Hence, $W_v(v,k)$ admits a unique Taylor series expansion around $v=0$. In Appendix~\ref{appen1}, the detailed proof of this statement has been documented.
 
Now taking $v\rightarrow0$ limit of $P_v(\omega,k,v)$ evaluated at its zero - $W_v^i(v,k)$ from Eq.~\eqref{show01},
\begin{equation}\label{show2}
\begin{split}
0=\lim_{v\rightarrow0}P_v\left(\gamma W_v^i(v,k),\gamma~k\right)&=\lim_{v\rightarrow0} P\left(\gamma W_v^i(v,k),\gamma~k\right) +\lim_{v\rightarrow0}~\sum_{r=1}^{r_{\rm max}} v^r Q^{(r)}\left(\gamma W_v^i(v,k),\gamma~k\right)~,\\
&=\lim_{v\rightarrow0} P\left(\gamma W_v^i(v,k),\gamma~k\right)= P\left(\left[\lim_{v\rightarrow0} W_v^i(v,k)\right],k\right)~.
\end{split}
\end{equation}
In other words, $\left[\lim_{v\rightarrow0}W_v^i(v,k)\right]$ must reduce to a zero of $P(\omega,k)$, i.e., a zero of the dispersion polynomial in LRF. But $P(\omega,k)$ has only $M<M_v$ number of zeroes. So, this is possible only if multiple distinct zeroes of $P_v(\omega,k,v)$ reduce to the same zero of $P(\omega,k)$ or have the same limit as $v\rightarrow 0$.

But we have shown in appendix~\ref{appen1} that these distinct zeroes of the boosted polynomial not only agree in the strict $v\rightarrow0$ limit but are also the same in a neighbourhood of $v=0$.
Now, all zeroes of $P_v(\omega,k,v)$ are smooth functions of $v$ as long as $|v|<1$, and we know that if two analytic functions agree in a neighbourhood, then they must be the same function. Therefore,  those zeroes of $P_v(\omega,k,v)$ that reduce to a single zero of $P(\omega,k)$ in the $v\rightarrow0$ limit must be the same functions at all $v$. But the effect of the boost cannot simply be an increase in the multiplicity of the roots of the LRF dispersion polynomial, as is clear from the factorization in equation \eqref{boostfactor}.
 
So it follows that the zeroes of $P_v(\omega,k,v)$ that have a finite limit as $v\rightarrow0$ are in one-to-one correspondence with the LRF modes. In other words, $P_v(\omega,k,v)$ must have $M_v-M$ number of zeroes that do not have a finite limit as $v\rightarrow 0$ and are called the spurious modes.\\

To summarize:
\begin{itemize}
\item Dispersion polynomials are polynomials in $\omega$ with coefficients that are again polynomials in $k$ (as expressed in \eqref{mostgen}).
\item The degree of the boosted polynomial $P_v(\omega,k,v)$  (denoted as $M_v$) will be larger than the degree of the dispersion polynomial $P(\omega,k)$ in LRF (denoted as $M$), whenever $P(\omega,k)$ has a term of the form $\left[(\text{some nonzero coefficient})\times(\omega^m k^n)\right]$, with $m<M$ but $(m+n)>M$. The maximum of such $(m+n)$ will be equal to $M_v$ = the number of zeroes in the boosted polynomial $P_v(\omega,k,v)$.
\item Among the $M_v$ number of zeroes of $P_v(\omega,k,v)$, there will be $M$ number of zeroes, which, in the limit of zero boost (limit $v\rightarrow 0$), will smoothly reduce to the $M$ zeroes of $P(\omega,k)$. These are the modes that have been explicitly worked out in the previous Section~\ref{nonspurious}.
\item The rest of $(M_v-M)$ zeroes of $P_v(\omega,k,v)$ will diverge in the $v\rightarrow 0$ limit. These are the spurious modes that simply do not exist in the LRF.
\end{itemize}
\subsection{Spurious modes and many-to-one mapping of the \texorpdfstring{$k$}{} plane: The issue with causality}

As explained in Section~\ref{general}, the spurious modes correspond to multiple solutions for $p$ from the first equation of \eqref{solpara2} (see the discussion around equation \eqref{boostsol}). Clearly, whether such multiple solutions exist or not entirely depends on the form of $W^i(k)$ or the mode in LRF.
In this subsection, we would like to show that whenever the form of $W^i(k)$ allows for such multiple solutions for the equation, $\gamma\left\{p-v~W^i(p)\right\}-k=0$ for every real $v<1$ and complex $k$, it violates the causality constraints given in Ref.~\cite{Heller:2022ejw}.

First, consider a case when $W^i(p)$ is a bounded function, i.e., $|W^i(p)|<B$ $\forall p$  where $B$ is a finite positive number. Here, if we take the $v\rightarrow 0$ limit on \eqref{solpara2}, we find,
$$\lim_{v\rightarrow 0~}k=\lim_{v\rightarrow 0} \gamma\left(p-v~W^i(p)\right) = p,\quad \lim_{v\rightarrow 0~}\omega = \lim_{v\rightarrow 0} \gamma\left(W^i(p)-v~p\right)= W^i(p)~.$$
In other words, if $W^i(p)$ is of finite modulus for all $p$, then the zeroes of $S_i$ and, hence, the modes of the boosted frame (see equation \eqref{boostfactor} and the discussion around it for the definition of $S_i$), in the zero boost limit, will smoothly reduce to a particular LRF mode.

On the other hand, we have already argued that if the factor $S_i$ has multiple zeroes for every $k$ and $v$, then some of them must be spurious modes, and therefore, the corresponding $\omega(k,v)$ must diverge in the $v\rightarrow 0$ limit (see the previous subsection).

From the above discussion, it follows that if spurious modes exist, then $W^i(p)$ must be some unbounded function of $p$.

Now suppose, $p= \bar p(v,k)$ is a solution to the first equation of \eqref{solpara2}. Then the expression for the corresponding zero of $P_v(\omega,k,v)$ has the following structure (from the second equation of \eqref{solpara2}),
\begin{equation}\label{eq:zero}
\begin{split}
\omega(v,k) =  \gamma\left\{W^i(\bar p)-v~\bar p\right\}~.
\end{split}
\end{equation}
If $\omega(v,k)$ is a spurious mode, then Eq.~\eqref{eq:zero} must admit that $\lim_{v\rightarrow 0}\omega(v,k)\rightarrow\infty$. In the following chain of arguments, we will demonstrate the conflict between the existence of spurious modes and the causality of the theory.

First, consider the case where $\lim_{v\rightarrow 0}\bar p(v,k) = \text{finite}=p_0$. Then, $\omega(v,k)$ to be a spurious mode in Eq.~\eqref{eq:zero}, $W^i(p_0)$ must diverge. But $W^i(p)$ is actually a mode in LRF, and according to Ref.~\cite{Heller:2022ejw} (Theorem 2 therein), no causal mode in LRF can have poles or essential singularities, i.e., it cannot diverge at any $p$ of finite modulus. 
So, for a spurious mode given in Eq.~\eqref{eq:zero} that respects causality, it claims that $p_0$ cannot be finite or $\lim_{v\rightarrow 0}\bar p(v,k)$  must diverge for all $k$.

Also, $W^i(p)$ must diverge separately for the existence of a spurious mode (as we argued before that the function $W^i(p)$ must be unbounded by any finite positive number for spurious modes to exist). So $W^i(p)$ should be such that its modulus diverges for large $p$ (and therefore at large $p_0$), since it cannot have poles or essential singularities, i.e., it cannot diverge at finite $p$.
So, for $\omega(v,k)$ in Eq.~\eqref{eq:zero} to be a spurious mode and require that causality be respected,
we must have (i) $W^i(p)$ diverging at large $p$, and (ii) $\lim_{v\rightarrow 0} \bar{p}(v,k)$ diverging for all $k$, simultaneously.
 
It follows that in the limit of small $v$, we should be able to find $\bar p(v,k)$ using the large $p$ expansion of $W^i(p)$.
Since, for a spurious mode, $\bar p(v,k)$ exists with a diverging zero-boost limit for all $k$, for simplicity, let us set $k=0$.
Let's also assume that $W^i(p)$ diverges at large $p$ as $C_1 ~p^\alpha, \text{where}~~\alpha>0$ and $C_1$ is some $p$ independent constant. It can be expressed as, \footnote{The power of the sub-leading terms ($\beta$) would also be fixed if $W^i(p)$ were maintained under the large-$k$ causality criteria, following \cite{Hoult:2023clg}.}
$$W^i(p) = C_1 ~p^\alpha + {\cal O}(p^{\alpha-\beta}),~~~\beta>0~.$$
Substituting this leading behavior into the first equation of \eqref{solpara2} with $k=0$ we find that,
\begin{align}\text{if} ~~\alpha\neq1,~~~&\bar p-v \left[ C_1 ~\bar p^\alpha + {\cal O}(\bar p^{\alpha-\beta})\right]=0~,\nonumber\\
~\Rightarrow~&\bar p = \left(1\over v~C_1\right)^{1\over\alpha-1}\left(1 + \text{terms vanishing at $v\rightarrow0$ limit} \right)~.
\end{align}
Now the above solution for $\bar p$ will diverge in the $v\rightarrow 0$ limit, only if $\alpha>1$.

For the case $\alpha=1$, the equation reduces to the following form,
$$(1 - v~C_1)\bar p = {\cal O}\left(\bar p^{1-\beta}\right)~~\Rightarrow~~\left( {1\over v}-C_1\right)={\cal O}\left(\bar p^{-\beta}\right)~.$$
In this case, a large $\bar p$ solution is possible only if ${1\over v}$ is very close to $C_1$, which is a finite number. Since at $v\rightarrow 0$, $\frac{1}{v}$ cannot be close to a finite number, the $\alpha=1$ case is excluded from the solution of $\bar{p}$.

In summary, the $v\rightarrow 0$ limit of $\bar p$ could diverge only if the LRF mode $W^i(p)$ diverges at large $p$ as $p^\alpha$ with $\alpha>1$.

In Ref.~\cite{Hoult:2023clg}, the authors have analyzed the allowed asymptotic (large momentum expansion) behavior of a mode in LRF, which is consistent with causality. It has been shown that for a causal theory, at large $p$ the mode $W^i(p)$ can diverge at most linearly (See Eq.(14) of \cite{Hoult:2023clg}, below we are quoting the leading terms of that equation),
$$W^i(p) = C_1 ~p +{C_2\over p} + {\cal O}(p^{-2}),~~~|C_1|<1~.$$
From the above discussion, we argue that such a mode can never generate a solution for $\bar p$ that diverges in the zero boost limit.
\\ \\
To summarize: \\ \\
The LRF mode $W^i(p)$ must diverge to generate a spurious mode. 
Consistency with causality demands that $W^i(p)$ cannot have any pole or essential singularity, implying that it cannot diverge at finite $p$,
but it allows $W^i(p)$ to diverge at large $p$. However, causality also requires $W^i(p)$ not to grow faster than linear, and the coefficient of the linear term must have a modulus less than one. Finally, we show that with this type of linear growth, it is never possible to find a parametric solution for the zeroes of $P_v(\omega,k,v)$ that diverges in the $v\rightarrow 0$ limit, as a spurious mode should do.

So we have shown explicitly that if the LRF spectrum is consistent with causality, then under boosts, the spurious modes will not be generated. In other words, whenever spurious modes are generated, i.e., whenever the number of modes is not conserved under boost transformation, some LRF mode(s) must violate the causality constraints.
\subsection{A simple example: Shear channel of relativistic Navier-Stokes equation}
So far, our discussion has been abstract. In this subsection, we shall apply the above discussion to the dispersion polynomial in the shear channel of the relativistic Navier-Stokes equation. This is a simple polynomial that is linear in $\omega$ and quadratic in $k$, and therefore violates the mode-preserving condition. It is also well-known that this is the mode that violates causality.

We shall first apply the boost transformation to the dispersion polynomial in the LRF. In this simple case, we could solve exactly for the zeroes of the boosted polynomial. Between the two modes, one turns out to be spurious, i.e., diverging at the $v\rightarrow 0$ limit. The other one in the zero-boost limit smoothly merges with the LRF mode.

Next, we analyse the parametric solution for the zeroes of the boosted polynomial as described in Eq.~\eqref{solpara2}. We clearly observe that the two-to-one map from the $p$ plane to the $k$ plane (since for every $k$, this is again a quadratic equation for $p$), leads to the same two zeroes of $P_v(\omega,k,v)$ that we already derived by exactly solving the boosted polynomial.

In LRF, the polynomial takes the following form,
\begin{equation}\label{eq:shear}
\begin{split}
P(\tilde \omega, \tilde k) &= \tilde \omega + i\eta ~\tilde k^2~,
\end{split}
\end{equation}
with $\eta$ as a constant parameter of the underlying theory.
Let $\omega$ and $k$ denote the frequency and the wavenumber in the boosted frame, and $P_v(\omega,k,v)$ is the dispersion polynomial for the same. Then we have the following relations,
\begin{equation}
   \tilde\omega = \gamma\left(\omega + v ~k\right)~,~~~~\tilde k = \gamma\left(k +v~\omega\right)~, 
\end{equation}
that follow the boosted polynomial to become,
\begin{equation}\label{eq:boost}
\begin{split}
P_v(\omega,k,v) =& ~P(\gamma\left(\omega +v ~k\right),\gamma\left(k +v~\omega\right))\nonumber\\
=&~\gamma\left(\omega + v ~k\right)+ i\eta ~\gamma^2\left(k +v~\omega\right)^2~.
\end{split}
\end{equation}
In this case, it is easy to solve for the zeroes of $P_v(\omega,k,v)$. The two zeroes are the following,
\begin{equation}\label{zeroeq}
\begin{split}
\omega_\pm(k)=\left(1\over 2i\eta v^2\gamma \right)\left[(-1 - 2i\eta v\gamma k) \pm\sqrt{1+4i\eta v\left(k\over\gamma\right)}\right]~.
\end{split}
\end{equation}
Let's study the $v\rightarrow 0$ limit of these two roots,
\begin{equation}\label{vzero}
\begin{split}
&\sqrt{1+4i\eta v\left(k\over\gamma\right)} = 1 + 2 i (\eta v) k -2 (i\eta v)^2 k^2 + {\cal O}(v^3)~.\\
\Rightarrow~&\omega_+(k) = -i\eta~k^2 + {\cal O}(v)~,\\
\Rightarrow~&\omega_-(k) = \left(1\over i\eta\right)\left[-{1\over v^2} -2i\left(\eta\over v\right) k + {\cal O}(1)\right]~.
\end{split}
\end{equation}
Clearly $\omega_-$ is the spurious mode that diverges in the $v\rightarrow 0$ limit, whereas $\omega_+$ smoothly merges with the LRF mode.

Now we shall derive these two roots using the multi-valued nature of the parametric solution.
The parametric solution for the zeroes of $P_v(\omega,k,v)$ has the following structure,
\begin{equation}\label{parasol}
\begin{split}
\omega = \gamma(-i\eta ~p^2- v~p),~~~k = \gamma(p+i\eta v~p^2)~.
\end{split}
\end{equation}
Here $p$ is the parameter. If we substitute equation \eqref{parasol} in the expression of $P_v(\omega,k,v)$, it identically vanishes.
From the second equation of \eqref{parasol}, we could see that for a fixed $k$, we could always find two solutions for $p$ as,
\begin{equation}\label{eqdouble}
\begin{split}
p_\pm(k) = \left(1\over 2i\eta v\right) \left[-1 \pm \sqrt{1 + 4 i\eta v\left(k\over\gamma\right)}\right]~.
\end{split}
\end{equation}
It is easy to check that around $v=0$, the roots $p_\pm$ admit the following expansion,
$$p_+(k,v) = k + {\cal O}(v),~~~~p_-(k,v) = \left(i\over\eta\right){1\over v} + {\cal O}(v^0)~.$$
Also, explicitly evaluating $\omega$ on $p_\pm$ we could generate the two zeroes of $P_v(\omega,k,v)$ from the first equation of \eqref{parasol} which readily gives,
$$\omega_+(k) =\omega(p_+),~~~\omega_-(k)=\omega(p_-)~.$$
Note, while using the parametric solution, we have not directly solved for the zeroes of $P_v(\omega,k,v)$, but instead used the LRF mode to generate the zeroes.

\section{Summary and outlook}\label{summary}

\subsection{Concluding remarks}

In this work, we have developed a general framework to study how the spectrum of linearized perturbations of a system around its equilibrium transforms across inertial frames connected by a Lorentz transformation. Considering the frequency of the perturbations $(\omega)$ as an infinite series in the momentum $(k)$, we provide here a precise mapping between the modes in two inertial frames connected by a Lorentz transformation. We find that, under certain conditions, one risks generating unphysical branches in the roots of the boosted dispersion relations, which we call the ``spurious modes". These spurious modes are necessarily divergent at the zero boost limit and lead to the violation of causality criteria in both the small and large wavenumber limits.

Focusing on hydrodynamics, considering $\omega$ as an infinite series in $k$, we first provide an exact mapping between the expansion coefficients in the LRF mode of the fluid and those in a Lorentz boosted frame with respect to the LRF, as long as all the boosted modes have a rest frame analogue. This gives us the liberty to estimate the boosted spectra (both hydrodynamic and non-hydrodynamic modes) solely from the knowledge of LRF modes without the need to solve the entire boosted polynomial. We emphasize here that this mapping is established for any general dispersion polynomial without referring to any particular theory. Thus, although discussed here in the setup of relativistic hydrodynamics, this framework can be useful in various other setups where the direct extraction of roots in a different inertial frame becomes extremely difficult. The ``$\gamma$-suppression" of expansion coefficients in the boosted frames discussed in this context also provides a hint towards understanding the interplay of stability and causality in the spatially homogeneous limit at ultra-high boosts. 

Next, we discuss how, at the level of the solutions of $\omega(k)$, the existence of spurious modes is a consequence of the multi-valued nature of the Lorentz transformation connecting the momenta in the LRF and the boosted frame. It is shown that from the knowledge of the $\omega$ and $k$ powers in the LRF dispersion polynomial, one can detect the presence of these spurious modes in the theory. We have also shown how the spurious modes must necessarily be divergent in the zero-boost limit. Trying to account for this divergent behavior in the small momentum limit leads to the existence of poles or essential singularities, which is causally prohibited by the analysis in \cite{Heller:2022ejw}. A similar attempt in the large momentum limit leads to either superluminal group velocities or a faster than linear $k$ growth of $\omega$, both of which violate the causality conditions discussed by \cite{Hoult:2023clg}. 

Thus, we demonstrate through a rigorous first-principle calculation why the existence of spurious modes necessarily leads to the violation of causality; hence, they act as good diagnostic probes to detect causality violations in a theory. Since one can uniquely determine a mode in the boosted frame from the information in the LRF, the generation of extra modes indicates that the dispersion polynomial in the boosted frame must be of a higher degree in $\omega$ than in the LRF, producing an excess number of roots. Therefore, the non-conservation of the number of modes in a theory necessarily renders it acausal. 

We illustrate the techniques developed in our work through two simple examples. The first example concerns a Maxwell-Cattaneo type diffusion equation. This serves as the stable-causal dispersion polynomial to demonstrate how one can exactly map the modes in the LRF to those in the boosted frame. The second example discusses the dispersion polynomial in the shear channel of the relativistic first-order Navier-Stokes equation, where we show how the multi-valued nature of the Lorentz transformation of momentum leads to spurious modes.

Altogether, the framework developed here presents a systematic approach to studying the spectrum of linearized perturbations in Lorentz boosted frames for a variety of effective theories where a dispersion polynomial can be defined. Although we have focused on relativistic hydrodynamics for some part of the discussion, it should be noted that similar dispersion spectra and causality issues can arise in many other scenarios, such as higher-derivative theories of gravity or in quantum or classical systems, thus indicating its applicability in a variety of other setups beyond hydrodynamics.

\subsection{Future directions and phenomenological implications}

Since the analysis presented here works exclusively with the solutions of the dispersion polynomials, without explicitly referring to the polynomials themselves, it can be applied in a variety of effective theory setups.

A first direction is the systematic classification of all effective theories that preserve the number of modes under Lorentz transformations. Combined with the techniques and results presented in \cite{Bhattacharyya:2024ohn}, these could be used to systematically construct the most general form of a theory with higher-derivative corrections that does not violate the causality criteria presented in \cite{Heller:2022ejw,Hoult:2023clg}. Such a systematic classification would, in principle, include theories with different underlying microscopic frameworks, such as kinetic theory or gravitational theory. Beyond relativistic hydrodynamics, where we would first like to test our case, such an analysis can also be used to diagnose issues with stability and causality in higher-derivative theories of gravity. 

A second possible extension could be to use the method presented in Section~\ref{nonspurious} to calculate the $\omega$ in a boosted frame, instead of explicitly extracting the roots from the equations of motion in a boosted frame. This may be applied to numerical relativistic hydrodynamics, where spurious roots often contaminate simulations at finite grid spacing or large flow velocities. Embedding our analytic mapping into numerical codes could help identify unphysical solutions, stabilise simulations, and guide the design of causal algorithms~\cite{Baier:2006gy,Adams:2006sv,Babichev:2007dw,Molnar:2009tx,Plumberg:2021bme,Chiu:2021muk,ExTrEMe:2023nhy,Hoshino:2024qun,Gavassino:2025bsn}.

A third application of our results lies in utilizing the ``$\gamma$-suppressed" nature of the mode-structures at small wavenumbers in the boosted frame for a multitude of purposes. Section~\ref{nonspurious} showed that the non-spurious boosted modes naturally organize themselves into series in $k/\gamma$, with higher-order terms increasingly suppressed at large boosts. Consequently, the higher-order terms of $\omega$ in a $k-$expansion become less significant with increasing boost. At near-luminal values of boost, this leads to all contributions in $\omega$ coming from only the leading order term, forcing all the sub-leading terms to vanish. Hence, as observed in \cite{Roy:2023apk}, at near-luminal values of boost, this leading term determines the physical constraints of this theory. While the reason for this behavior is clear with the understanding of $\gamma-$suppression presented here, it would be beneficial to have a rigorous analytical derivation to show how exactly a point on the complex-$k$ plane gets mapped to a point in the large momenta region following a Lorentz-transformation of near-luminal boost, along with taking a $k\to 0$ limit. 

Apart from these mathematical intricacies, the result of $\gamma-$suppression also leads to interesting physical consequences. Firstly, it naturally regulates the behavior of higher-order terms in the $k-$expansion of $\omega$. These higher-order terms are often the source of instabilities in certain hydrodynamic formulations. This also explains the often-observed empirical fact that boosted-frame analyses of relativistic hydrodynamic theories (especially in the context of numerical simulations) tend to exhibit improved stability and causal behavior. Secondly, this aligns with the observations in \cite{Basar:2024qxd,Bhambure:2024gnf,Bhambure:2024axa} that the dynamics of the long-wavelength perturbations in a highly boosted inertial frame are primarily governed by only a few of the leading-order terms in the $k-$expansion of $\omega$. This again indicates that the use of results in the near-luminally boosted frames can act as an effective probe of the short-wavelength limit of the theory, without directly venturing out of the long-wavelength hydrodynamic limit itself.

Further interesting extensions of conceptual interest lie in exploring whether the many-to-one mapping on the complex-$k$ plane, which leads to spurious modes, has a geometric interpretation. Since spurious modes appear precisely when the map ceases to be one-to-one, studying its singularities may reveal deeper structural constraints on effective field theories with derivative expansions.

Beyond formal consistency, the results have several direct implications for phenomenology--particularly for systems where relativistic flows and dissipative effects coexist, such as heavy-ion collisions, astrophysical plasmas, and condensed matter analogues~\cite{Rezzolla:2013dea,Arslandok:2023utm,Berges:2025izb}: 
\begin{enumerate}
    \item Hydrodynamic modelling of heavy-ion collisions: Boosted backgrounds are ubiquitous in ultra-relativistic nuclear collisions, where longitudinal flow velocities approach $v \approx 1$. Our results show that first-order or improperly truncated hydrodynamic theories inevitably generate spurious boosted roots, even if they appear stable in the LRF. This explains why certain dissipative formulations fail in highly boosted regions such as the fragmentation zone or early-time longitudinal expansion. The criteria derived here may provide a diagnostic tool for determining which hydrodynamic schemes remain physically valid across the full kinematic domain.
    \item Interpretation of numerical instabilities: Many hydrodynamic codes exhibit instabilities at large flow velocities or at the boundary of applicability, often attributed to numerical artifacts~\cite{Pandya:2021ief}. Our analysis indicates that these instabilities might arise due to spurious boosted branches leading to spurious oscillations in the computational domain. 
    Thus, the spectral analysis developed here provides a method to distinguish numerical artifacts from genuine physical breakdowns of the effective theory.
    \item Signals for the breakdown of hydrodynamic applicability: In regimes where gradients become large or where the theory is pushed beyond its causal domain~\cite{Fantini:2025gnm}, the appearance of boosted spurious modes provides a clean and operational indicator that hydrodynamics has ceased to be reliable. This may have implications for interpreting observables in small collision systems or in the far-from-equilibrium regime~\cite{Noronha:2024dtq,Grosse-Oetringhaus:2024bwr}.
\end{enumerate}
\section*{Acknowledgements}
S.B. would like to acknowledge Shiraz Minwalla for valuable discussions and useful insights.
S.M. acknowledges the Department of Atomic Energy, India, Grant No.~RIN4001 for financial support. 
R.S. is supported by a postdoctoral fellowship from the West University of Timișoara, Romania.  
S.R. would like to express his gratitude to the people of India for their steady and generous support for research in basic sciences.

\appendix

\section{Proof for uniqueness of boosted modes: analyzing \texorpdfstring{$v$}{} derivatives of \texorpdfstring{$P_v(\omega,k,v)$}{} and its zeroes}
\label{appen1}
Suppose $\omega = W_v(v,k)$ is a zero of $P_v(\omega,k,v)$, which in the zero-boost limit smoothly merges with the LRF mode $W(\tilde{k})$. Hence, it follows that, 
\begin{equation}\label{show1}
\begin{split}
0&=P_v( W_v(v,k),k)= P(\gamma W_v(v,k),\gamma k)+\sum_{r=1}^{r_{\rm max}} v^r Q^{(r)}(\gamma W_v(v,k),\gamma k)~.
\end{split}
\end{equation}
Now, we shall take the $v\rightarrow 0$ limit on the $n^{th}$ derivative of Eq.~\eqref{show1}. We note that, 
\begin{equation}\label{intst}
\begin{split}
&\lim_{v\rightarrow 0}{d^n\over dv^n}\sum_{r=1}^{r_{\rm max}} v^r Q^{(r)}(\gamma W_v(v,k),\gamma k)\\
 = &\lim_{v\rightarrow 0}\sum_{m=0}^n \left[n!\over (n-m)!m!\right]\sum_{r=1}^{r_{\rm max}}\left(d^m v^r\over dv^m\right)\left(d^{n-m}\over dv^{n-m}\right)Q^{(r)}(\gamma W_v(v,k),\gamma k)~,\\
  = &\lim_{v\rightarrow 0}\sum_{r=1}^{{\rm min}(n,r_{\rm max})}~~\sum_{m=0}^{{\rm min}(n,r)} \left[n!\over (n-m)!m!\right]\left(d^m v^r\over dv^m\right)\left(d^{n-m}\over dv^{n-m}\right)Q^{(r)}(\gamma W_v(v,k),\gamma k)~,\\
  = &\sum_{r=1}^{{\rm min}(n,r_{\rm max})}~~\sum_{m=0}^{{\rm min}(n,r)} \left[n!\over (n-m)!m!\right]\left[r!\over (r-m)!\right]\delta_{r,m}\left[\lim_{v\rightarrow 0}\left(d^{n-m}\over dv^{n-m}\right)Q^{(r)}(\gamma W_v(v,k),\gamma k)\right]~,\\
    = &\sum_{r=1}^{{\rm min}(n,r_{\rm max})}\left[n!\over (n-r)!\right]\left[\lim_{v\rightarrow 0}\left(d^{n-r}\over dv^{n-r}\right)Q^{(r)}(\gamma W_v(v,k),\gamma k)\right]~.
    \end{split}
\end{equation}
Now, the $n^{th}$ order derivative of any nested function $F(Y(x))$ could be expressed as,
\begin{equation}\label{rule0}
{d^n\over dx^n}F(Y(x))= \left[d^n Y(x)\over dx^n\right]\left[dF(z)\over dz\right]\bigg|_{z = Y(x)}  + F_{\rm lower}(x)~,
\end{equation}
where $F_{\rm lower}$ contains at most $(n-1)^{th}$ derivatives of the function $Y(x)$.

Using Eq.~\eqref{rule0} we find,
\begin{equation}\label{rule1}
\begin{split}
{d^n\over dv^n}P(\gamma W_v(v,k),\gamma k)&= \left[\left(\partial^n\over \partial v^n\right)\bigg(\gamma W_v(v,k)\bigg)\right]\left(\partial P(\omega,\gamma k)\over \partial\omega\right)\bigg|_{\omega = \gamma W_v(v,k)}  + P_{\rm lower}(v)~,\\
{d^n\over dv^n}Q^{(r)}(\gamma W_v(v,k),\gamma k)&= \left[\left(\partial^n\over \partial v^n\right)\bigg(\gamma W_v(v,k)\bigg)\right]\left(\partial Q^{(r)}(\omega,\gamma k)\over \partial\omega\right)\bigg|_{\omega = \gamma W_v(v,k)}  + Q^{(r)}_{\rm lower}(v)~,
\end{split}
\end{equation}
where both $ P_{\rm lower}$ and $Q^{(r)}_{\rm lower}$ contain at most $(n-1)^{th}$ derivatives of the function $\gamma W_v(v,k)$.
Substituting \eqref{intst}, \eqref{rule0} and \eqref{rule1} in the zero boost limit of the $n^{th}$ derivative of Eq.~\eqref{show1} we find,
\begin{equation}\label{show3}
\begin{split}
0=~&\lim_{v\rightarrow0}\left(d^n\over dv^n\right) P_v(W_v(v,k),k)\\
=~& \lim_{v\rightarrow0}\left(d^n\over dv^n\right)\left[P(\gamma W_v(v,k),\gamma k)\right]+\lim_{v\rightarrow0}\left(d^n\over dv^n\right)\left[\sum_{r=1}^{r_{\rm max}} v^r Q^{(r)}(\gamma W_v(v,k),\gamma k)\right]~,\\
=~& \lim_{v\rightarrow0}\left[\left(\partial^n\over \partial v^n\right)\bigg(\gamma W_v(v,k)\bigg)\right]\left(\partial P(\omega,{\gamma}k)\over \partial\omega\right)\bigg|_{\omega = \gamma W_v(v,k)}  +\lim_{v\rightarrow 0} P_{\rm lower}(v)\\
&~~~~~~~~~~~~~~+\sum_{r=1}^{{\rm min}(n,r_{\rm max})}\left\{n!\over (n-r)!\right\}\left[\lim_{v\rightarrow 0}\left(d^{n-r}\over dv^{n-r}\right)Q^{(r)}(\gamma W_v(v,k),\gamma k)\right]~,\\
=~&\left\{\partial P(\omega,{\tilde{k}})\over \partial\omega\right\}\bigg|_{\omega ={W(\tilde{k})}}~ \lim_{v\rightarrow0}\left[\left(\partial^n\over \partial v^n\right)\bigg(\gamma W_v(v,k)\bigg)\right] \\
&~+\sum_{r=1}^{{\rm min}(n,r_{\rm max})}\left[n!\over (n-r)!\right]\left[\lim_{v\rightarrow 0}\left(d^{n-r}\over dv^{n-r}\right)Q^{(r)}(\gamma W_v(v,k),\gamma k)\right] +\lim_{v\rightarrow 0} P_{\rm lower}(v)~.
\end{split}
\end{equation}
If we assume that
$\left(\partial P(\omega,{\tilde{k}})\over\partial\omega\right)\bigg|_{\omega =W(\tilde{k})}$ is non-zero, 
then equation \eqref{show3} is an algebraic equation that uniquely determines the $n^{th}$ derivative of $\gamma W_v(v,k)$ (and therefore $W_v(v,k)$) at the $v\rightarrow 0$ limit in terms of the lower order $v$ derivatives evaluated at the same limit. So we can use equation \eqref{show3} to recursively determine $v$ derivatives of $W_v(v,k)$ up to all orders at the $v\rightarrow 0$ limit. Also, recursively, we could see that if all $v$ derivatives of $W_v(v,k)$ up to the $(n-1)^{th}$ order are finite in the zero-boost limit, then the $n^{th}$ order derivative will also be finite. Further, we have assumed that the zeroth derivative of $W_v(v,k)$ (i.e., the function itself) has a smooth and finite zero boost limit to some unique LRF mode $W(\tilde{k})$. Therefore, using equation \eqref{show3} we could uniquely determine $W_v(v,k)$ in an infinite expansion around $v=0$.
In other words, if we know the zero boost limit of $W_v(v,k)$ is finite and equal to an LRF mode, then we know $W_v(v,k)$ uniquely in the neighborhood of $v=0$, and it is finite.
\bibliographystyle{JHEP}
\bibliography{ref}
\end{document}